\documentclass[]{elsarticle}
\usepackage{algorithm,algorithmic}

\newcommand\T{\rule{0pt}{2.6ex}}
\newcommand\B{\rule[-1.2ex]{0pt}{0pt}}
\usepackage{graphicx,color,subfigure,colordvi,epsfig,amssymb,amsmath,amsfonts,verbatim,url,textcomp,xspace}
\newcommand{\DynaChanAl}[1]{\textit{DynaChannAl}}

\date{}

\begin{document}

\title{DynaChannAl: Dynamic Channel Allocation with Minimal End-to-end
  Delay for Wireless Sensor Networks}

\author[jhucs]{JeongGil Ko\corref{cor1}}
\ead{jgko@cs.jhu.edu}

\author[jhucs]{Amitabh Mishra}
\ead{amitabh@cs.jhu.edu}

\cortext[cor1]{Corresponding author}

\address[jhucs]{Department of Computer Science, Johns Hopkins University, Baltimore, MD 21218}

\begin{abstract}
  With recent advances in wireless communication, networking, and low
  power sensor technology, wireless sensor network (WSN) systems have
  begun to take significant roles in various applications ranging from
  environmental sensing to mobile healthcare sensing. While some WSN
  applications only require a limited amount of bandwidth, new
  emerging applications operate with a noticeably large amount of data
  transfers. One way to deal with such applications is to maximize the
  available capacity by utilizing the use of multiple wireless
  channels.  This work proposes \DynaChanAl~, a distributed dynamic
  wireless channel algorithm with the goal of effectively distributing
  nodes on multiple wireless channels in WSN systems.  Specifically,
  \DynaChanAl~ targets applications where mobile nodes connect to a
  pre-existing wireless backbone and takes the expected end-to-end
  queuing delay as its core metric. We use the link quality indicator
  (LQI) values provided by IEEE 802.15.4 radios white-list potential
  links with good link quality and evaluate such links with the
  aggregated packet transmission latency at each hop. Our approach is
  useful for applications that require minimal end-to-end delay (i.e.,
  healthcare applications). \DynaChanAl~ is a light weight and highly
  adoptable scheme that can be easily incorporated with various
  pre-developed components and pre-deployed applications. We evaluate
  \DynaChanAl~ in on a 45 node WSN testbed. As the first study to
  consider end-to-end latency as the core metric for channel
  allocation in WSN systems, the experimental results indicate that
  \DynaChanAl~ successfully distributes multiple (mobile) source nodes
  on different wireless channels and enables the nodes to select
  wireless channel and links that can minimize the end-to-end latency.
\end{abstract}


\maketitle

\section{Introduction}
\label{sec:intro}

The recent advances in both sensing and wireless communication
technology have motivated researchers to study the applicability of
wireless sensor networks (WSNs) in various applications.  Most WSN
platforms use IEEE 802.15.4-based~\cite{ieee802.15.4} radios to
achieve extremely low power consumption. While the simple design of
IEEE 802.15.4 offers significantly lower power consumption compared to
other wireless standards such as IEEE 802.11~\cite{ieee802.11}, their
bandwidth is significantly limited to only 250 Kbps in an ideal
environment. We note that in practice when using widely used software
and hardware platforms, even this low rate is impossible to
achieve~\cite{petrova2006performance}. While early WSN applications
that have low data rate requirements~\cite{SSC+06} can properly
operate even with such strict wireless capacity limitations, various
emerging applications require significantly higher data rates to
support the increasing amount of sensing data~\cite{medisn,
  racnet}. One way to provide such high data rate applications with
sufficient wireless channel capacity is the \emph{smart} use of
\emph{multiple} wireless channels. Doing so extends the usable
bandwidth within a system and since the Zigbee alliance~\cite{zigbee}
defines 16 different channels for its systems in the 2.4 GHz band, we
can significantly increase the capacity of a WSN system by
accommodating a smart channel allocation scheme.

Previous work that propose channel allocation techniques for WSNs
mainly focus on balancing the number of nodes on each channel with the
assumption that all nodes generate the same amount of
traffic~\cite{WSH+08,ZHTHS06}.  Also, most of the proposed schemes
either target networks that consist of only stationary nodes or they
perform channel allocation only during the initial phase of the
deployment~\cite{WKZ+09}. Such characteristics make previously
proposed schemes less appropriate for applications with dynamic
traffic patterns which is common in WSN deployments \cite{KoGT09}.
Furthermore, given that in many WSN applications the sensing or
communication devices are mobile and that different devices can have
different data rate requirements, a dynamic channel allocation
technique should be used to utilize the use of multiple wireless
channels efficiently. Moreover, some applications such as medical
sensing applications~\cite{medisn,alarmnet} require the data to be
received at the gateway with minimal latency. Therefore, for such
applications, the end-to-end delay should act as a core metric that
determines which channel that a source node should operate on. Finally
given that WSN systems incorporate many different protocols and
schemes to optimize their performance for a specific application, the
channel allocation scheme should be easily adaptable with minimal
communication and memory overhead.

With such issues in mind, we propose \DynaChanAl~ a distributed
dynamic channel allocation scheme with the goal of effectively
utilizing multiple wireless channels while minimizing the end-to-end
latency of packets from mobile WSN nodes. As discussed above, since
many WSN applications require soft real-time data delivery, we take
the goal of minimizing end-to-end latency as \DynaChanAl~'s primary
goal. By minimizing the end-to-end delay, we target to see fairness in
packet delivery performance between multiple wireless channels as
well. Our scheme also makes use of the link quality indicator (LQI)
values reported by IEEE 802.15.4-based radios to whitelist or
blacklist potential links with respect to their wireless conditions
before they are selected for further monitoring (e.g., latency
measuring). \DynaChanAl~ targets to benefit the performance of WSN
systems that consist of a wireless mesh backbone network rooted at the
gateway and multiple mobile source nodes that associate to such
backbone network for delivering their data to the gateway using
multihop connections. To our knowledge, this work is the first to
propose the use of end-to-end latency as a core metric in the subject
of channel allocation for WSN systems.

Specifically in this work we make the following contributions.
\textbf{(1)} We propose \DynaChanAl~, a light weight distributed
dynamic channel allocation scheme for WSNs that takes estimated
end-to-end latency as the core metric for determining the wireless
channel conditions.
\textbf{(2)} We quantify the effectiveness and show the feasibility of
wireless channel allocation techniques for IEEE 802.15.4 radio based
WSN systems using a simple empirical study with real mote-class
devices.  \textbf{(3)} We validate the effectiveness of \DynaChanAl~
in real testbed environments to show that it is effective in
minimizing the end-to-end latency and distributing multiple nodes to
different wireless channels in real WSN systems despite being
extremely light weight.

The paper is organized as follows. In Section~\ref{sec:relwork} we
introduce existing literature related to wireless channel allocation
in the area of WSNs and discuss how the previously proposed techniques
are not appropriate for the application scenarios of our interest.
Next we introduce the results obtained from our empirical studies to
show the feasibility of channel switching in WSNs then discuss about
our metrics and propose our scheme, \DynaChanAl~ in
Section~\ref{sec:metric}. Our evaluation results obtained from the
testbed experiments are presented in Section~\ref{sec:eval} and we
discuss about some interesting aspects of \DynaChanAl~ in
Section~\ref{sec::discuss}. Finally, we conclude the paper by
introducing some potential future work (Section~\ref{sec:future}) and
with a conclusion in Section~\ref{sec:sum}.

\section{Related Work}
\label{sec:relwork}

Until now, most of the work in channel allocation for wireless sensor
networks (WSNs) have focused on evenly dividing the number of nodes on
each channel without considering the amount of traffic that each node
generates~\cite{WSH+08,ZHTHS06}.  These work hold the assumption that
the traffic generated by each sensor node will remain constant over
time and be equal for all nodes. However, we argue that WSNs can
introduce dynamic traffic patterns for various reasons. First, because
many WSN systems are event-driven~\cite{He04}, nodes can easily have
bursty traffic patterns. Such bursty traffic can quickly degrade the
performance of a channel allocation method when the nodes are
distributed on multiple channels in a static manner. Second, WSN
applications introduced recently can be mobile. Such applications have
dynamic traffic patterns due to the mobility of end-user
devices~\cite{medisn}. Also, such applications can support multiple
types of sensors and therefore, generate different amount of traffic
on each sensor.

To address such issues, some work have made attempts to perform
channel allocation based on the amount of traffic each node
generates. Wu et al.~\cite{WKZ+09}, consider traffic aware allocation
by computing the estimated traffic in a channel for proper channel
allocation to sensor nodes. However, their scheme is only performed in
the initial stage of the system deployment and therefore, cannot be
effective to tolerate dynamic channel environments.

Despite such previous work related to channel allocation for WSNs, one
main drawback of such schemes is that these schemes are not flexible
and therefore are not widely used. Specifically, the previous work
mentioned above tried to propose a comprehensive solution for the
entire system while most WSN systems are designed as a combination of
multiple pre-existing components~\cite{medisn}.  We argue that the
channel allocation scheme presented in this work, \DynaChanAl~, is
flexible enough to accommodate and coexist with protocols in different
layers of the network stack.  We present some examples in
Section~\ref{sec::discuss}.

We also note that previous work on channel allocation in WSNs such
as~\cite{chowdhury2009channel,salameh2007adaptive} discuss thoroughly
about the analytical aspects of WSN system's wireless channel
allocation.  However their evaluation on the proposed protocols are
only shown using simulation or analytical results which only partially
represent real wireless environments. This work is one of the few work
that propose a channel allocation scheme for WSNs and evaluate the
scheme with real devices on a testbed environment.

One example of previous work that has been evaluated in realistic
testbed environments is the work proposed by Le et al.~\cite{LHA08}.
Here, the authors proposed a MAC protocol for multi-channel systems
which targets general WSN applications. Their scheme is a control
theory based approach where the channel switching metric is directly
related to the channel accessibility (e.g., the number of CCA failures
that a node encounters). They optimize their protocol so that the
amount of cross channel communication and channel switching
fluctuation is minimized. While we see this as a notable approach with
valid evaluation methods, one drawback of their scheme is that the
scheme does not consider latency as a decision metric. We argue that
given the numerous emerging sensing applications that require
soft-real time data delivery, maintaining a maximum end-to-end latency
threshold and balancing the latency on multiple wireless channels
should be taken in consideration when nodes select between multiple
wireless channels.  Works proposed by He et
al. \cite{he2003speed,he2004aida} deal with wireless sensor networks
that have latency constraints but their schemes mostly focus on the
routing side of the system while \DynaChanAl~ discusses about channel
allocation techniques for WSNs.  Specifically, the SPEED protocol
\cite{he2003speed} requires that node to be aware of their physical
location which is not practical for mobile nodes and AIDA
\cite{he2004aida} only focuses on maximizing the utilization for a
single channel, both for making routing decisions.  To our knowledge
\DynaChanAl~ is the first work that focuses on channel allocation for
WSNs while taking the expected end-to-end delay as a core decision
metric.

\section{\DynaChanAl~} 
\label{sec:metric}

\begin{figure}[t] \centering
\includegraphics[width=0.7\linewidth]{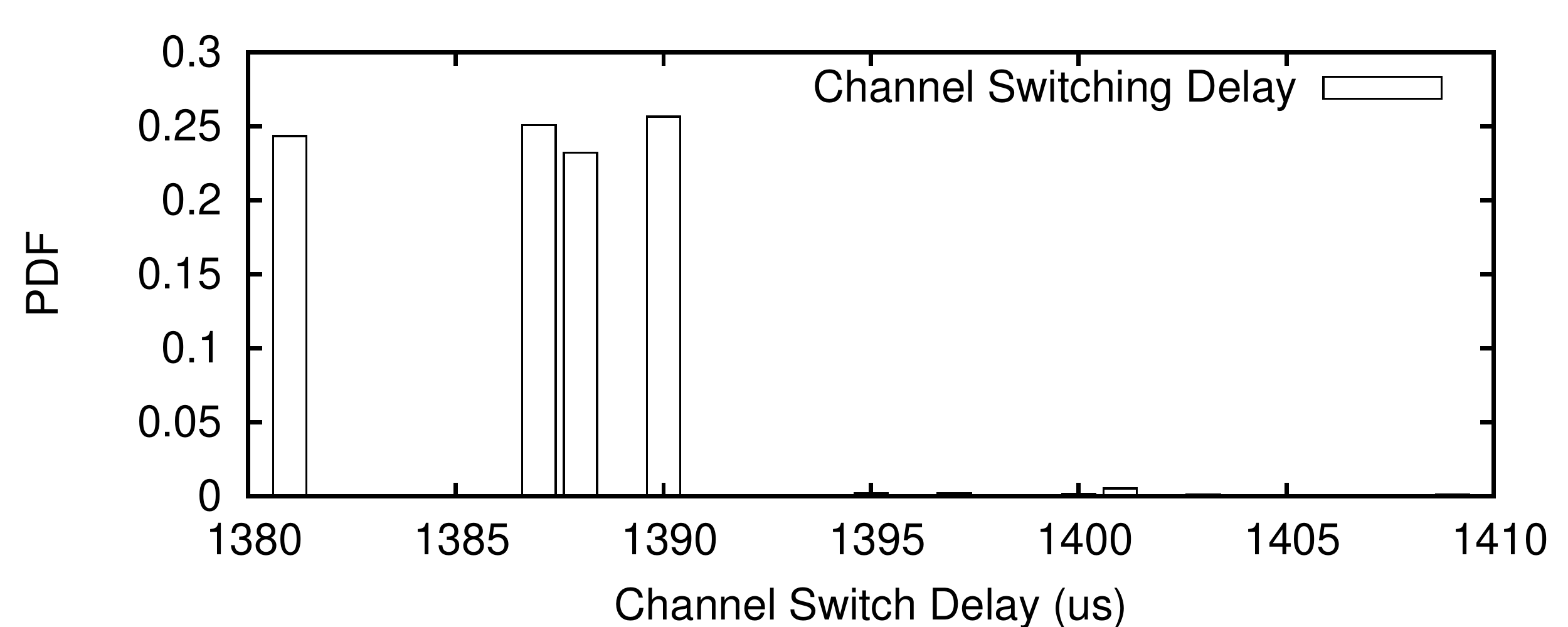}
\includegraphics[width=0.7\linewidth]{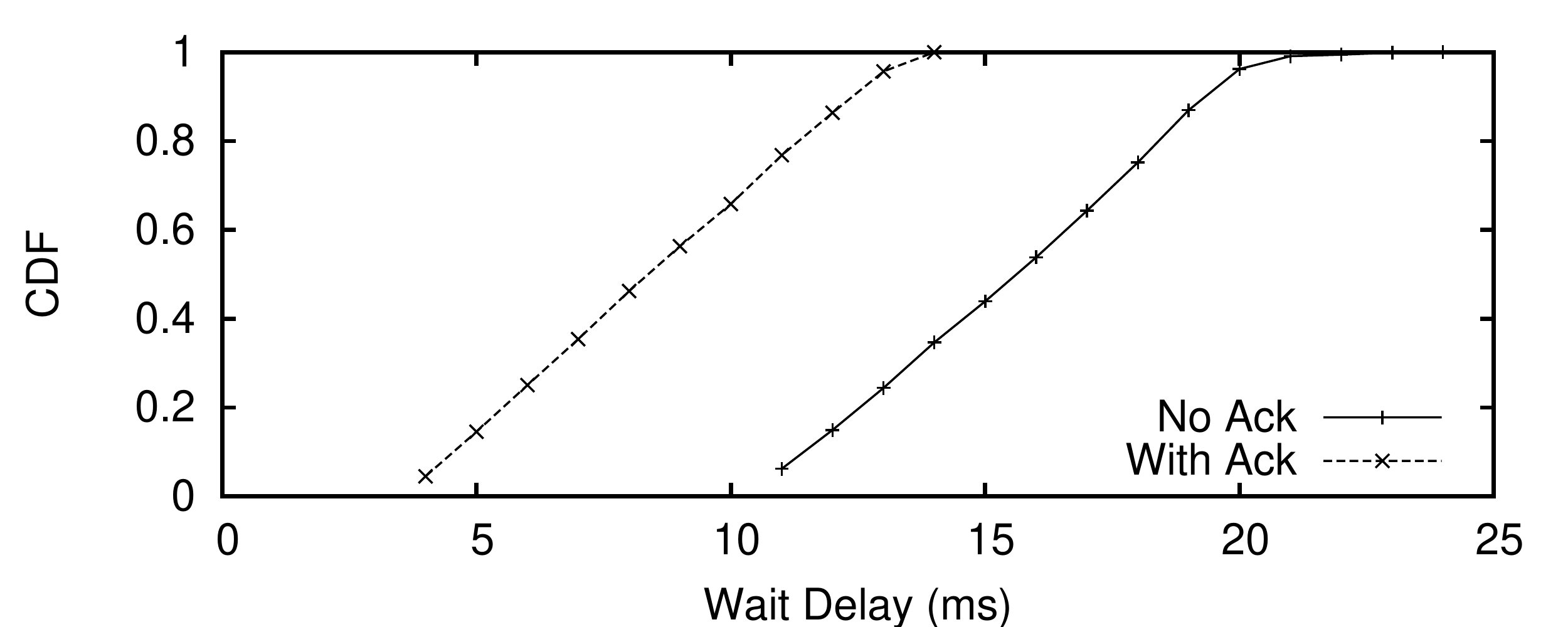}
\caption{PDF of the channel switch delay (top) and CDF of time
  required to finalize a send event with and without the existence of
  acknowledgment packets (bottom). Both tests are done with CC2420
  radios on a TMote Sky platform with TinyOS 2.1.}
\label{fig:delay} 
\end{figure}

\begin{figure}[t] \centering
\includegraphics[width=0.7\linewidth]{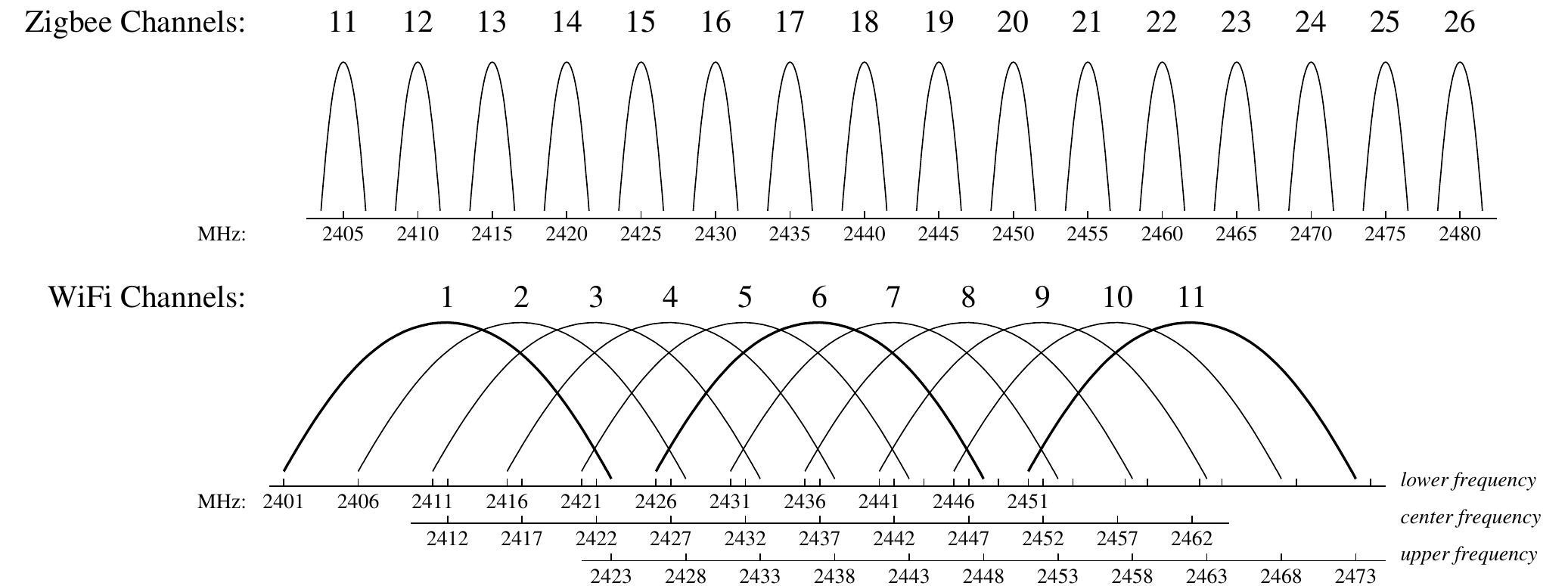}
\caption{Diagram of the 16 channels allocated for Zigbee traffic in the
2.4 GHz band. The bottom figure shows how the 802.11 frequencies are
distributed.} \label{fig:channel} \end{figure}

\subsection{Feasibility of Channel Switching in WSNs}

\label{sec:empirical}

We first show the results obtained from a simple empirical study to
confirm that the overhead of channel switching is minimal and
therefore, channel switching is an attractive way to maximize the
performance for WSN systems. We note that the main overhead of
switching between multiple wireless channels is the delay
overhead. The two main delay overheads that arise due to channel
switching are \textbf{(1)} the delay required to physically ``switch''
the wireless channel of the radio and \textbf{(2)} the delay for
``seeking'' the channels to determine the link quality with a
potential next hop node. We define the two types of delays as {\it
channel switching delay} and {\it channel seeking delay},
respectively. Our empirical study is performed with CC2420 radio based
Tmote Sky devices using a TinyOS 2.x based software platform. We show
the results of the two types of delay overheads in
Figure~\ref{fig:delay}. We note that we take at least 1000 samples for
each test case.

The top part of Figure~\ref{fig:delay} shows a PDF of how long it
takes to actually ``switch'' the channel in our experimental settings
(i.e., the channel switching delay).  The results indicate that most
of the channel switching delays take less than 1.4 ms. Given that the
number of usable wireless channels for IEEE 802.15.4 devices is 16
(see the Zigbee plot in Figure~\ref{fig:channel}), changing the
wireless channels from among all possible channels (i.e., worst case)
can be done within $\sim$23 ms. This short amount of time is tolerable
given that even for WSN applications that require relatively high data
rates, the data generation intervals are slower than 100
ms~\cite{medisn}.

The bottom plot in Figure~\ref{fig:delay} shows the CDFs of two
different test cases. In the first experiment, a transmitter sends
periodic packets with acknowledgment requests but fails in receiving
an acknowledgment packet from its destination node. Therefore the
transmitter waits for the maximum amount of time until it detects that
the packet transmission was not successful (i.e, no Ack). The second
experiment (i.e., with Ack), shows results from an experiment where an
acknowledgment frame is immediately received at the transmitter after
the packet is sent. In this case, instead of waiting for the maximum
amount of time, the transmission process terminates as soon as the
acknowledgment is received.  Again, given that we have a maximum of 16
channels to seek (see Figure~\ref{fig:channel}), in the worst case,
the maximum idle waiting delay is $\sim 350 ms$ when no
acknowledgments are received on any of the 16 channels. In other
words, a source node can recognize the existence of different nodes in
its communication range (potentially nodes that can forward its
packets) operating on any of the 16 channels within a maximum time of
$\sim$350 ms.  We note that when acknowledgment frames are received on
a subset of these channels, this channel seeking delay can be much
lower. Therefore, a scheme where nodes actively probe the channel for
potential relay nodes can be an effective way of determining their
existence and link qualities on each wireless channel\footnote{On each
  wireless channel, multiple potential next hop nodes exist and the
  link qualities are different for each connection. We use the term of
  selecting a ``link'' and selecting a ``channel'' interchangeably in
  the work.}. We will describe in detail later in this work how this
observation is used in \DynaChanAl~.

To conclude, even in the worst case where a node must seek all 16
channels and wait for the maximum amount of time on each channel, the
channel seeking delay plus the channel switching delay is less than
$\sim 400 ms$. Such small delay can keep the idle listening time low
and can be used as evidence that channel switching schemes are
desirable for resource constraint WSN systems.

\subsection{Defining the Target Scenario and System Metrics}
\label{sec:def_metric}

\begin{figure}[t] \centering
\includegraphics[width=0.5\linewidth]{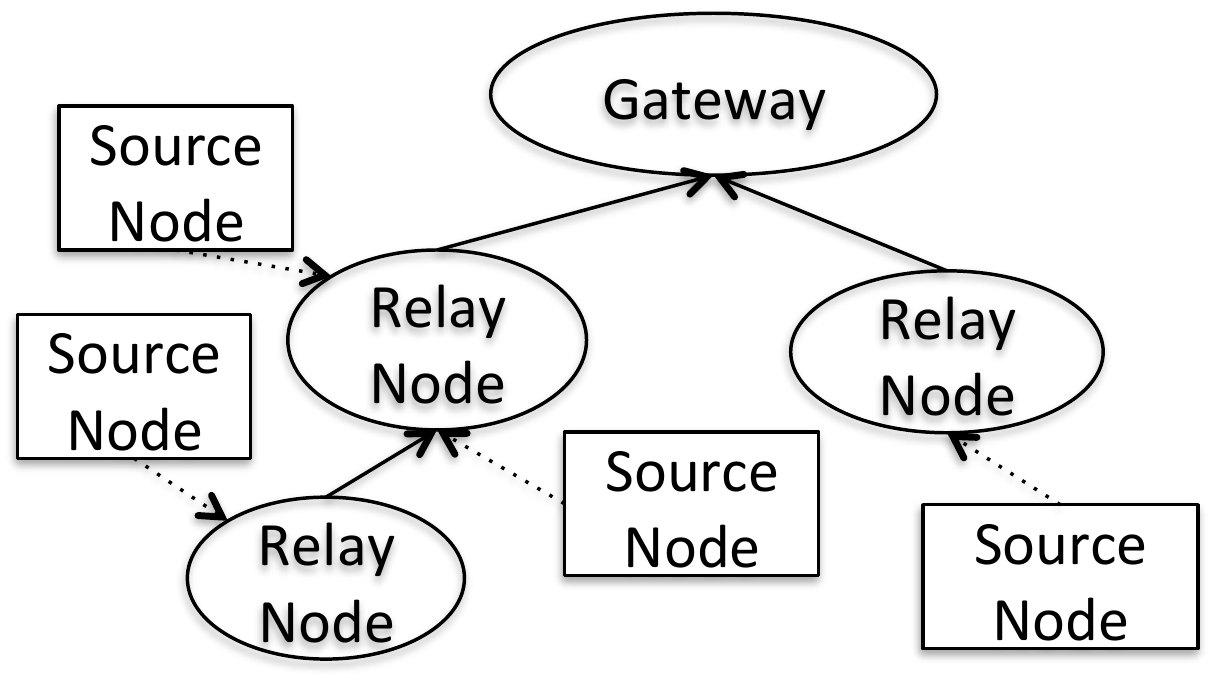}
\caption{Diagram of \DynaChanAl~'s target scenario. \DynaChanAl~
  targets applications that have a wireless backbone for data
  relaying. Data packets are originated from mobile source nodes that
  connect to the backbone as leaf nodes. Specifically, the backbone
  collects data from mobile source nodes (via single hop
  communication) and forwards this data to a gateway using a multihop
  tree network. In the diagram the solid edges with the oval vertices
  represent the backbone network and the mobile source nodes are
  presented in rectangled vertices and dotted edges.}
\label{fig:app-scenario} \end{figure}

To design an algorithm that efficiently utilizes multiple channels for
WSN systems, we must first define metrics that enables a node to
effectively compare the performance of potentially usable wireless
channels.  To suit our target application~\cite{medisn} (i.e., a
medical sensing application), we take the expected end-to-end latency
as the core metric. Additionally, we also use the link quality
indicator (LQI), offered by IEEE 802.15.4 based radios, to determine
the quality of potential links that connect a mobile source node and
the next hop relay node (more details on why we say ``relay'' node is
presented in the following subsection). Overall, since good quality
links with high LQI will lead to less first hop retransmissions, thus
reduce the first hop latency, the LQI metric is used as an indicator
of how good the \emph{local} network conditions are, while the
expected end-to-end latency reflects the network conditions on an
\emph{end-to-end} perspective. We use the LQI metric over the received
signal strength indicator (RSSI) given that it is well known through
previous literature that the \emph{aggregate LQI} is a more accurate
estimate of the link quality than RSSI \cite{PSC05,SL06}. The
following subsections discuss about our target usage scenario of our
proposed channel allocation scheme and the two metrics (e.g.,
end-to-end latency and LQI) in further detail.

\subsubsection{Target Scenario} 

Our proposed channel allocation scheme targets WSN applications that
operate on a multi-tier network. Specifically, \DynaChanAl~ targets
systems where a backbone network already exists with the goal of
forwarding the data from mobile source nodes to a common destination
(e.g., a gateway node).  The benefits of maintaining a backbone
network rather than a fully ad-hoc single-tier networks have been
shown in our previous work \cite{KoGT09}. Examples of such systems
include previously proposed medical sensing applications such as
MEDiSN~\cite{medisn} and AlarmNet~\cite{alarmnet}. In these
applications the mobile source nodes associate themselves to one or
more relay nodes and act as leaf nodes of the tree (or backbone)
network. For such networks, the role of \DynaChanAl~ is to balance the
traffic load and latency of the WSN system by controlling the wireless
channel of individual mobile source nodes in a distributed manner. In
this work we assume that wireless backbone infrastructure exists on
multiple wireless channels that IEEE 802.15.4 systems can operate
on. A diagram of our target scenario's network hierarchy is
illustrated in Figure~\ref{fig:app-scenario}.

\subsubsection{Expected End-to-End Queuing Delay} 

We define our first and main metric, the expected end-to-end queuing
delay as a metric that provides the source node with a global view of
the wireless channel conditions within path to the
gateway/destination. This metric will not only provide information on
how a packet will encounter the local channel environment but also
show how the wireless channel conditions will be throughout its path
to the packet's final destination in an aggregated way. We define the
expected end-to-end queuing delay at node $i$ as
Equation~\ref{eq:queueDelay}.

\begin{equation} 
D_i^q = \sum_{j \in P_i} d_j^q
\label{eq:queueDelay}
\end{equation}

where, $d_j^q$ is the local queuing delay at node $j$ and $P_i$ is the
set of nodes on the path from node $i$ to the root of the tree (i.e.,
first relay node to the gateway in Figure~\ref{fig:app-scenario}).
Note that set $P_i$ includes node $i$ (i.e., the source node) as
well. Therefore, $D_i^q$ can be simply computed by all the nodes in
the network in a distributed fashion using the following procedures.

\begin{itemize}

\item{Each node $i$ maintains an Exponentially Weighted Moving Average
    (EWMA) with $\alpha = 0.5$ of its own queuing delay ($d_i^q$) for
    packets that it transmits. This information is propagated to its
    children. In our implementations the measured queuing delay has a
    granularity of milliseconds and therefore, can be measured with a
    simple timer implementation in common software platforms.}

\item{The child node $k$ that receives this delay information from a
    parent node computes its own $D_k^q$ using
    Equation~\ref{eq:queueDelay}.}

\item{Finally, this updated end-to-end queuing delay is propagated to
    $k$'s children as well. This propagation can be done using
    explicit control packets or by piggybacking the information on
    data packets.}

\item{If a data packet requires retransmissions due to the lack of
    acknowledgments (when requested) or the lack of available
    bandwidth on node $k$'s local wireless channel, the additional
    time that the packet spends in the node's queue is aggregated to
    $D_k^q$, and therefore implicitly includes the local congestion
    level of the network.}

\end{itemize}

The procedure described above assures that all nodes in the network
can easily compute its own $D^q$ and quickly update the queuing delay
changes when dynamic traffic conditions occur (due to the fast
propagation of delay values). Here, we make the assumption that the
channel conditions will stay constant for $\Delta t$ seconds where
$\Delta t$ is the time needed for a node to decide the wireless
channel it will transmit its packet on. This is a valid assumption for
two reasons.  First, we note that in most cases the delay values will
be fairly stable. In our target applications mobile source nodes are
not used to forward other source nodes' data. Therefore, the local
queuing delays of other source nodes will not affect a source node's
delay measurements. Knowing that the $D^q$ from the potential next hop
node will only be affected by the latency variation introduced from
the \emph{stationary} backbone network, it is intuitively true that
the stationary nodes will have less variance in queuing delay
measurements than those of the mobile nodes. Second, as we pointed out
in Section~\ref{sec:empirical} that the time needed for channel
switching and seeking is small. Such small delay will lead to a small
value of $\Delta t$, making the assumption practically valid. If a
list of potentially usable wireless channels is given as an input
during the initialization stage or as a parameter in run-time, $\Delta
t$ can be further minimized.

We note that all delay values used in $D_i^q$ can be accurately
measured using resource constraint mote-class devices, by using a
simple timer implementation and four bytes of memory space for each
packet in its queue (when using 32 bit timestamps). Such simple
computation and piggybacking techniques (mentioned in the third item
above) makes the use of the $D_i^q$ measurements, feasible with
minimal memory and communication overhead.

\subsubsection{Link Quality Indicator (LQI)}
\label{sec:metric_lqi}

\begin{figure}[t] \centering
\includegraphics[width=0.5\linewidth]{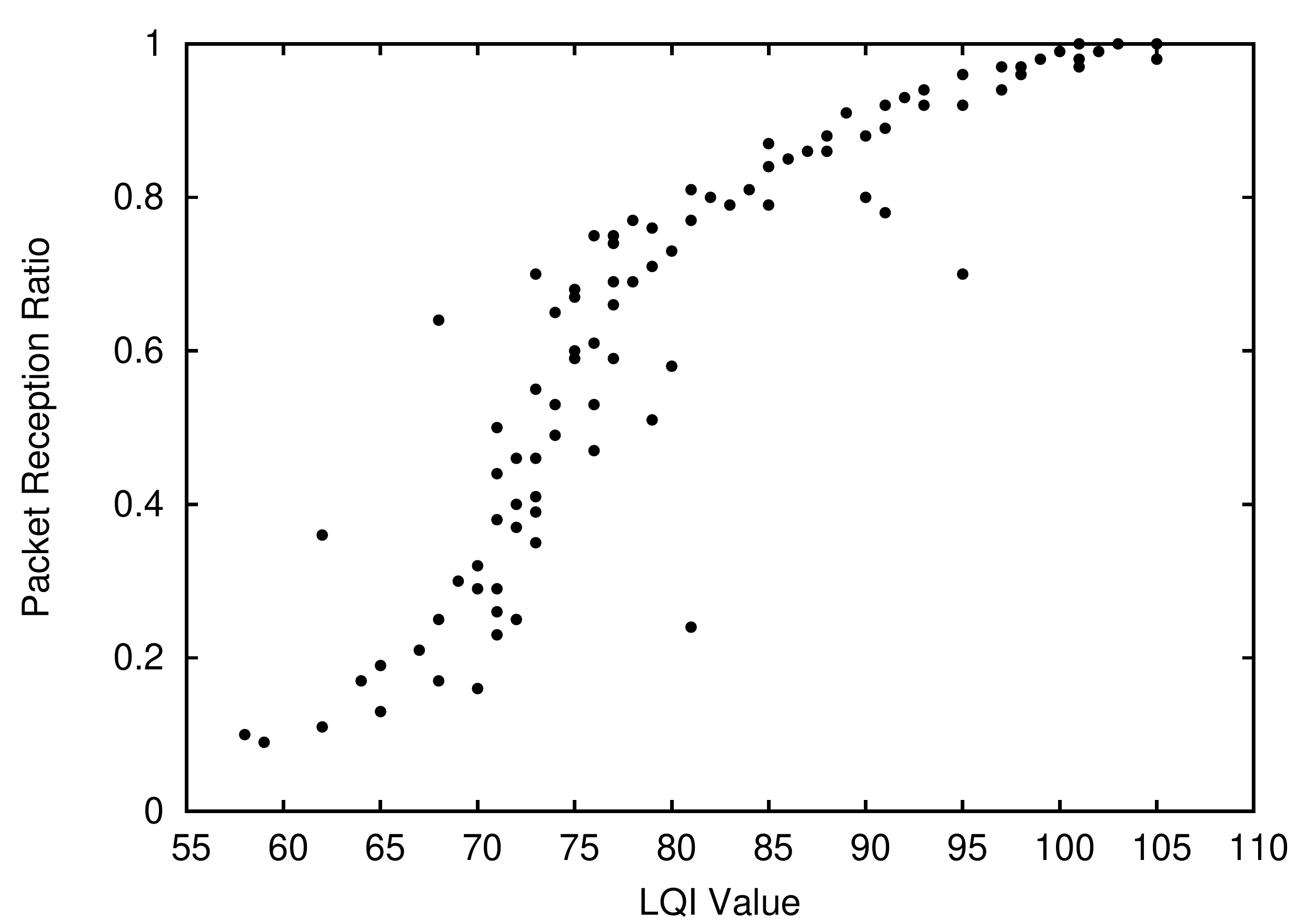}
\caption{Average LQI values with respect to the packet reception ratio
  of each link. The y-axis represents the packet reception ratio (PRR)
  collected from each link averaged over every minute.}
\label{fig:lqi-empirical} \end{figure}

While the measurement and estimation of end-to-end delay is a way of
indicating the end-to-end network conditions, there is also a need to
maintain a metric that represents the wireless channel conditions in a
local network's perspective as well. Typically in a local
\emph{single-hop}'s perspective, the quality of the links determine
the expected number of packet retransmissions and thus can represent
the expected single hop latency (i.e., more retransmissions lead to
longer queuing delays). Therefore, we are interested in defining a
metric that represents the conditions of each individual link that a
source node can use to minimize the number of retransmissions. The
link quality indicator (LQI) is a metric that is specific to IEEE
802.15.4-based radios \cite{ieee802.15.4}. LQIs indicate the link
quality between the sender node and the receiver node pair on a per
packet basis. While the IEEE 802.15.4 standard does not specify a
method to compute the LQI value, the widely used TI/Chipcon CC2420
radio \cite{cc2420} defines LQI as a representation of the chip error
rate.  Given that the CC2420 radio is used in popular WSN mote
platforms (TelosB and MicaZ), we use this implementation of the LQI as
our third metric.

The LQI values can be collected without any additional timing or
memory overhead when using TinyOS 2.x as the software platform.  As
previous work shows \cite{PSC05}, high aggregate LQI values can be an
effective indicator of good quality links. We show similar results in
Figure~\ref{fig:lqi-empirical}, where we present the relationship
between average LQI values and the per minute packet reception ratio
(PRR) collected from Tmote Sky devices on our testbed. While we cannot
imply precise wireless channel condition estimates, we can notice that
LQI can provide an indication of \emph{good}, \emph{average} and
\emph{poor} quality links. We define an aggregate LQI metric, computed
for each sender-receiver pair, in Equation~\ref{eq:LQI}.

\begin{equation}
LQI_i = \frac{\sum_{k=0}^{n} LQI^k_i}{n}
\label{eq:LQI} 
\end{equation}

Here, $LQI_i^n$ represents the aggregate LQI of the link between node
$i$ and its parent for the $n$th packet. While the accuracy of $LQI_i$
partially relates to the number of packet exchanges, we use findings
from previous work~\cite{PSC05} as evidence that even a small number
of samples (i.e., as low as one, due to the small variance in LQI
values when the link conditions are good) can roughly classify the
link quality. Again, while with a small number of samples it is
difficult to accurately determine the estimated PRR but can at least
roughly classify the quality of a wireless link in three regions,
e.g., \emph{good}, \emph{fair}, and \emph{poor}. Results in previous
work \cite{SL06} support this idea and also provide evidence that a
higher $n$ (e.g., $LQI_i$) ``will'' indeed provide higher accuracy and
precision in link quality estimation. Therefore, we use $LQI_i$ to
monitor the link conditions in a long-term scale and use a small $i$
when roughly classifying link qualities when we quickly seek among
different channels. We will describe how we use $LQI_i$ for each case
with greater detail in Section~\ref{sec:algo}.  Furthermore, we note
that the observations from Figure~\ref{fig:lqi-empirical} show that we
can classify links with $LQI > 85$ as good links ($PRR > 80\%$), $75 <
LQI < 85$ as fair links ($50\% < PRR < 80\%$) and $LQI < 75$ as poor
quality links ($PRR < 50\%$).

\subsection{Selecting Channels} 
\label{sec:algo}

With the metrics defined in Section~\ref{sec:def_metric}, we describe
the specific procedures of \DynaChanAl~ in greater detail. Overall,
\DynaChanAl~ is a fully distributed scheme that consists of a channel
seeking phase and a channel monitoring phase. \DynaChanAl~ is a simple
and intuitive scheme with the goal of showing that even such a simple
channel allocation scheme can effectively distribute mobile WSN nodes
in multiple channels with respect to the goal of minimizing the
end-to-end latency. The following subsections will discuss about the
role of the two phases and their transitions.

\subsubsection{Channel Seeking Phase}
\label{sec:channel-seek}

When a source node joins the network, the channel seeking phase is
initiated. The channel seeking phase is the period when a source node
actively probes all possible wireless channels (or a subset when the
active channels are known) to determine which link (i.e., next hop
node) on which channel would be its best choice for its current
location (i.e., different physical locations can have different
performance on each wireless channel/link) and time instance. In this
phase, the mobile source nodes send probe packets that advertise its
presence on the channel and starts to classify each potential link
with the information provided by the reply messages that relay nodes
(i.e., the nodes on the backbone network; see Figure
\ref{fig:app-scenario}) send back. The results in Section
\ref{sec:empirical} can be used as evidence that this phase can be
done quickly, even if we seek all possible 16 channels. This short
duration is desirable given that during this period, the source node
is unable to transmit its data packets nor put its radio to sleep to
conserve energy.  Specifically, in the channel seeking phase mobile
source nodes perform the procedures presented in Algorithm
\ref{algo:channel-seek}.

\begin{algorithm}[t]
\caption{Channel Seeking Phase Procedure}
\label{algo:channel-seek}
  \begin{algorithmic}[1]
    \STATE Source node $i$ enters channel seeking phase and broadcasts
    a probe with its network ID $i$ on wireless channel $C$.

    \STATE Each relay node $j$ that receives this probe message
    responds with a reply packet that contains its most up-to-date
    expected $D_i^q$ measurement and network ID $j$.

    \STATE Source node $i$ receives the reply packet.

    \STATE Determines quality of link using the observations in
    Section~\ref{sec:metric_lqi}. Save the link quality classification
    for the link between node $i$ and $j$ (i.e., good / fair / poor).

    \STATE Save $D_i^q$ from the reply message and associate with
    current node $j$.

    \STATE Repeat steps 3-5 for all reply messages received from
    different nodes on a channel.

    \STATE $C \leftarrow C+1$. 

    \STATE If $C$ is a valid channel, repeat from step 1.

  \end{algorithmic}

\end{algorithm}

As a result, at the end of the channel seeking phase the source node
should have a view of the link quality for each potential relay node.
With this information, the source node first tries to select the node
with the \emph{smallest} expected delay measurements that is
classified to have a \emph{good} quality link connection. If no
\emph{good} quality connections exist, the node sequentially searches
for \emph{fair} and \emph{poor} quality links with respect to the
expected delay measurements to finally select the best possible
wireless channel and associates to a relay node on that specific
channel. 

We note that the size of the table where we store the link and delay
related information is at maximum proportional to the number of
potential next hop relay nodes that are within the source node's
communication range, therefore, storage is not a major constraint in
our scheme. Also note that while asymmetries in link quality is common
for low power wireless networks it mostly happens for links with low
quality links in the transitional region \cite{asymmetry}. Therefore,
by selecting links with good or fair quality and blacklisting poor
quality links to be in the lowest priority, \DynaChanAl~ is minimally
affected by link asymmetries despite using LQI measurements from a
single direction to infer the link quality of the other direction.

\subsubsection{Channel Monitoring Phase}
\label{sec:channel-monitoring}

\begin{figure}[t] \centering
\includegraphics[width=0.7\linewidth]{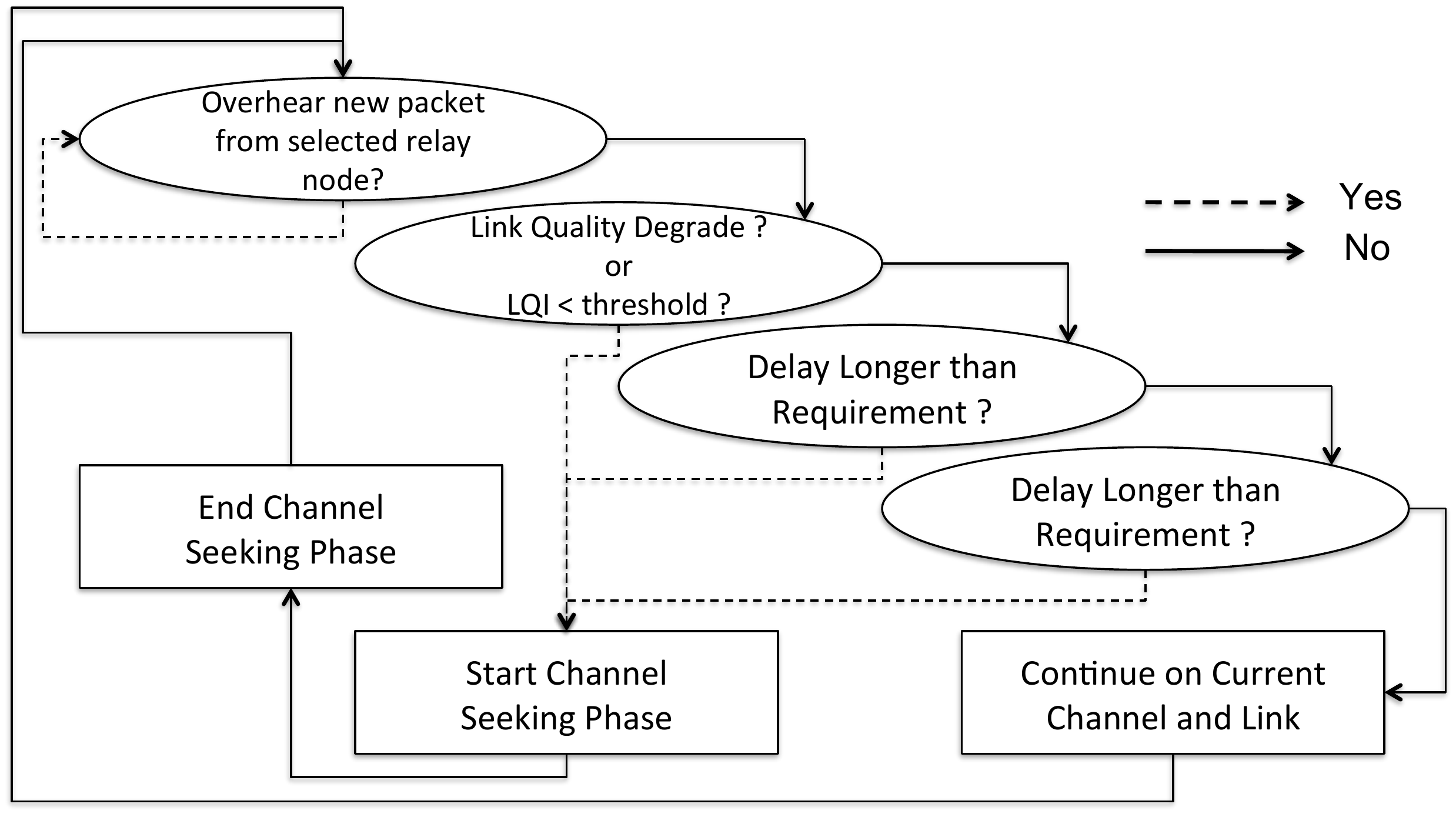} 
\caption{Decision tree diagram of the channel monitoring phase and its
  interaction with the channel seeking phase. At each channel
  monitoring phase the $LQI_i$ and $D_i^q$ values are recursively
  evaluated using the values that are stored at the end of the
  previous channel seeking phase. The channel seeking phase is
  re-invoked When the channel monitoring phase determines that the
  channel quality has degraded (e.g., in terms of LQI or delay) more
  than a pre-specified threshold.}
\label{fig:monitor-diagram} 
\end{figure}

The outcome of the channel seeking phase is the best wireless channel
and relay node that a source node can select in its ``current''
physical location and time instance. However, due to the potential
mobility and the dynamic traffic patterns of the source nodes that our
target applications introduce, we need a way to continuously monitor
and re-evaluate the selected wireless channel conditions. This
monitoring is to assure that the performance of the selected relay
node and wireless channel are still the best, or at least close to
what the source node expected when they were initially selected. This
leads to proposing an additional phase that monitors the current
status of the wireless channel and link conditions.

With this purpose, we start a \emph{channel monitoring phase} as soon
as the channel seeking phase terminates. In the channel monitoring
phase, the source node performs the operations shown in
Algorithm~\ref{algo:channel-monitor}.

\begin{algorithm}[t]
\caption{Channel Monitoring Phase Procedure}
\label{algo:channel-monitor}
  \begin{algorithmic}[1]

    \REQUIRE Source node ID $i$

    \STATE Set $\tau_D$ and $\tau_{LQI}$ in percentiles.

    \STATE Set application specific end-to-end delay requirement.

    \STATE $D_{channelInit} \leftarrow D_i^q$

    \STATE $LQI_{channelInit} \leftarrow LQI_i$

    \STATE $LQI_{classify} \leftarrow classification~of~selected~link$

    \STATE Node $i$ overhears messages sent by its associated relay
    node $j$ and collects the updated $D_j^q$ along with the new LQI
    ($LQI_{new}$) from the packets. Compute updated $LQI_i$ with
    $LQI_{new}$.

    \IF {$LQI_i \ge (LQI_{channelInit} \times \tau_{lqi})$}

    \IF {$New~LQI~classification \ge LQI_{classify}$}

    \IF {$D_i^q < delay~requirement$}

    \IF {$D_i^q \ge (D_{channelInit} \times \tau_D)$}

    \STATE Continue on current channel

    \ELSE

    \STATE Begin Channel Seeking Phase

    \ENDIF

    \ELSE

    \STATE Begin Channel Seeking Phase

    \ENDIF

    \ELSE

    \STATE Begin Channel Seeking Phase

    \ENDIF

    \ELSE

    \STATE Begin Channel Seeking Phase

    \ENDIF

    \STATE When new packets are overheard, repeat procedure from step 6.

  \end{algorithmic}
\end{algorithm}

In the channel monitoring phase operations, the LQI value comparisons
assure that the one hop link quality of the current connection has not
degraded significantly compared to when the link was initially
selected and the delay measurement comparison confirms that both the
absolute and relative values of the current end-to-end delay estimates
are still tolerable. We summarize this channel monitoring phase as a
decision tree in Figure~\ref{fig:monitor-diagram}.


We note that this monitoring phase that consists of four simple
comparisons (as specified in lines 7-23 in Algorithm
\ref{algo:channel-monitor}) is extremely light weight with minimal
memory constraints and takes minimal processing time since there are
no significant computation needed in the process. Such characteristics
of the scheme make \DynaChanAl~ desirable for resource constraint WSN
systems.



'\section{Evaluation}
\label{sec:eval}

\begin{figure}[t] \centering
\includegraphics[width=0.7\linewidth]{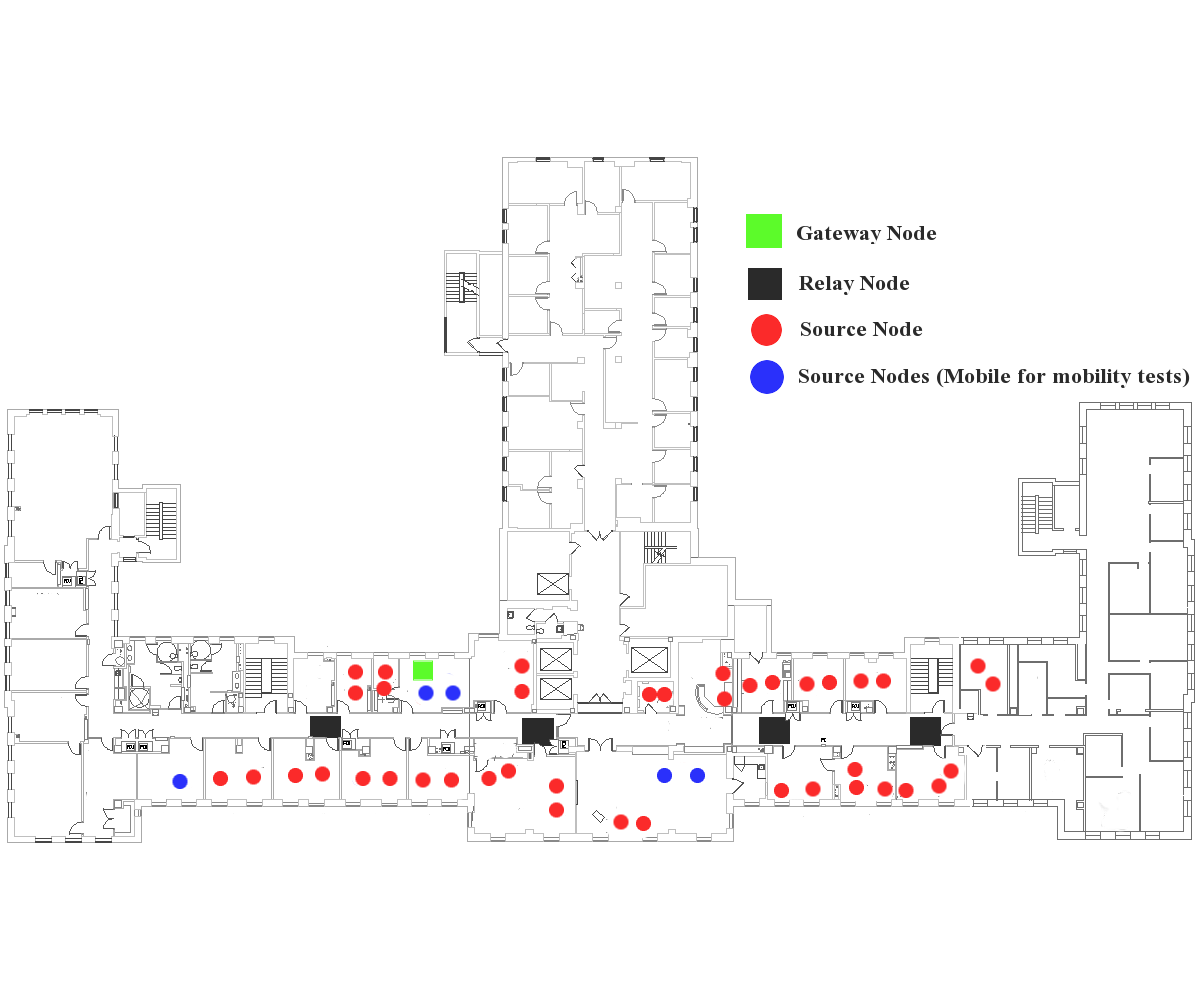}
\caption{Pictorial overview of testbed and the positions of relay
  devices. The light green colored square is the gateway and the dark
  squares represent the positions of the relay nodes. For each
  position two nodes (one for each channel 25 and 26) were
  positioned. The blue circles indicate the source nodes that were
  used as mobile devices in the mobility tests.}
\label{fig:testbed} 
\end{figure}

\begin{figure}[t] \centering
\includegraphics[width=0.49\linewidth]{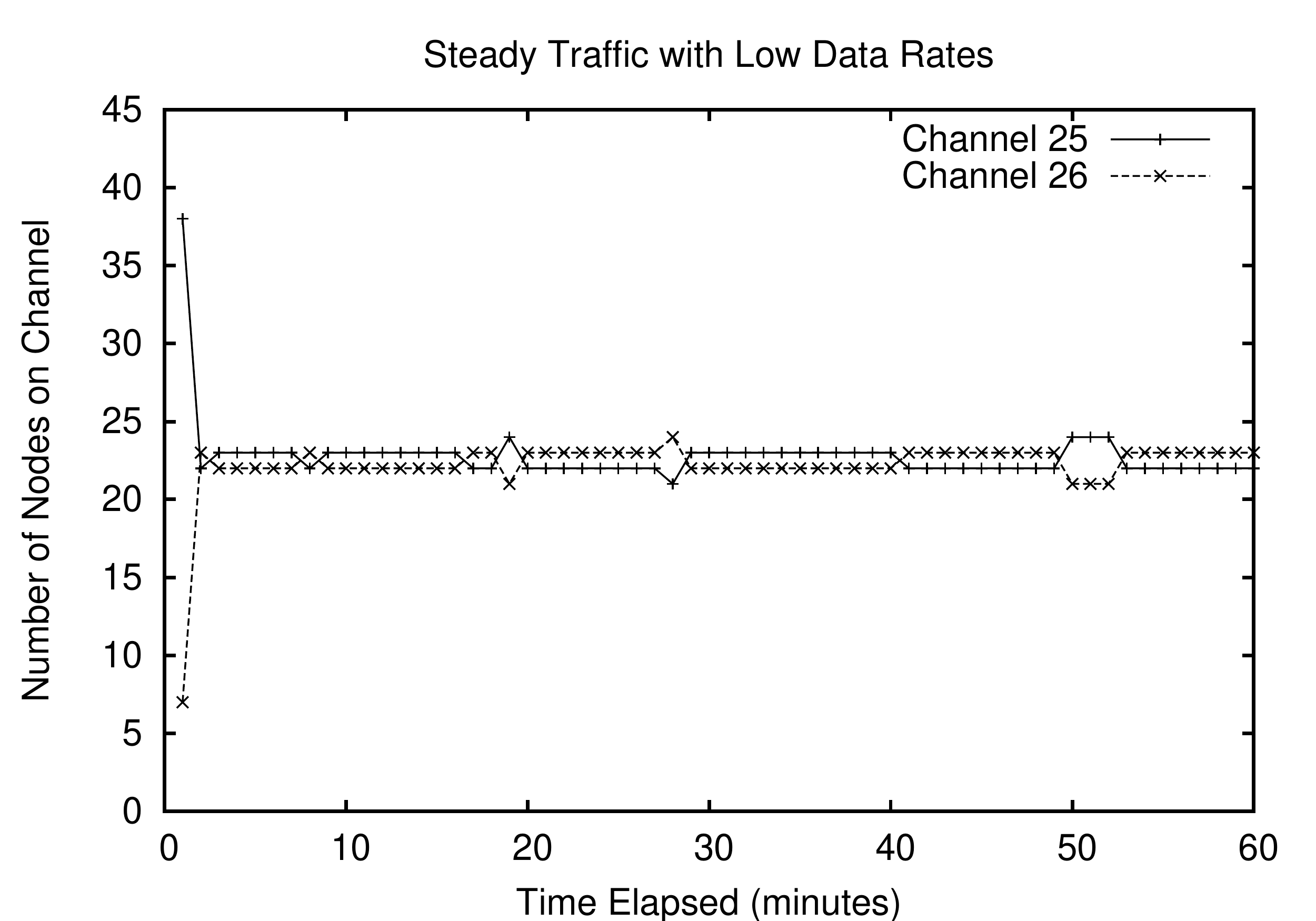}
\includegraphics[width=0.49\linewidth]{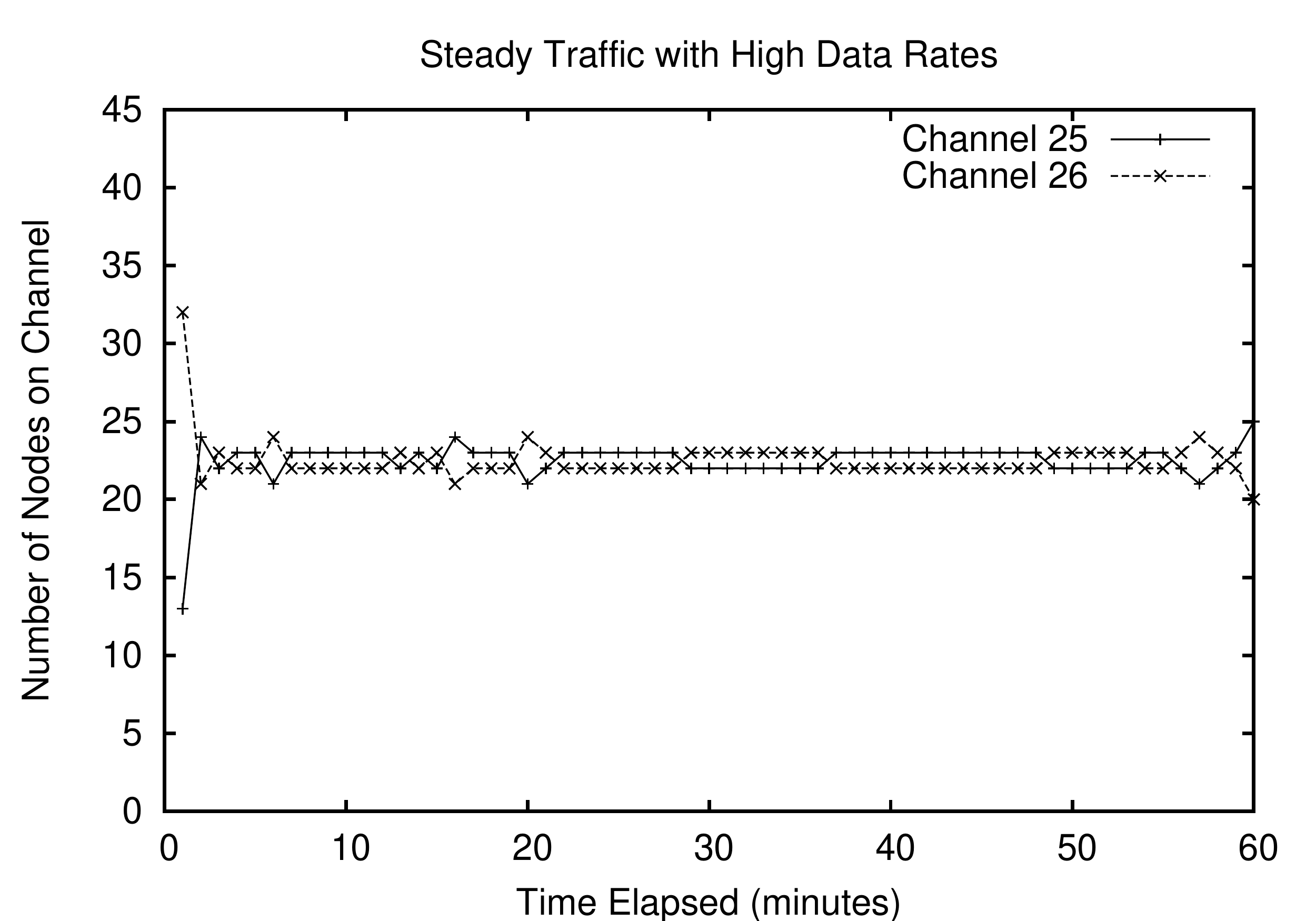}
\caption{Distribution of 45 nodes on each of the two operating
  channels over time with steady-periodic traffic. The left plot shows
  the case when packets are generated at an interval of 1024 ms and
  the right plot shows the case for traffic with 128 ms interval at
  each source node. The distribution of nodes on the two channels is
  uneven in the beginning of the experiment due to the fact that the
  relay nodes cannot provide the source nodes with sufficient
  information related to the expected end-to-end channel
  delay. However shortly after the experiment begins, nodes
  self-distribute themselves evenly on channels 25 and 26.  We can see
  that the distribution of the nodes are fair between the two wireless
  channels.}
\label{fig:channel-select-steady} 
\end{figure}

\begin{figure}[t] \centering
\includegraphics[width=0.49\linewidth]{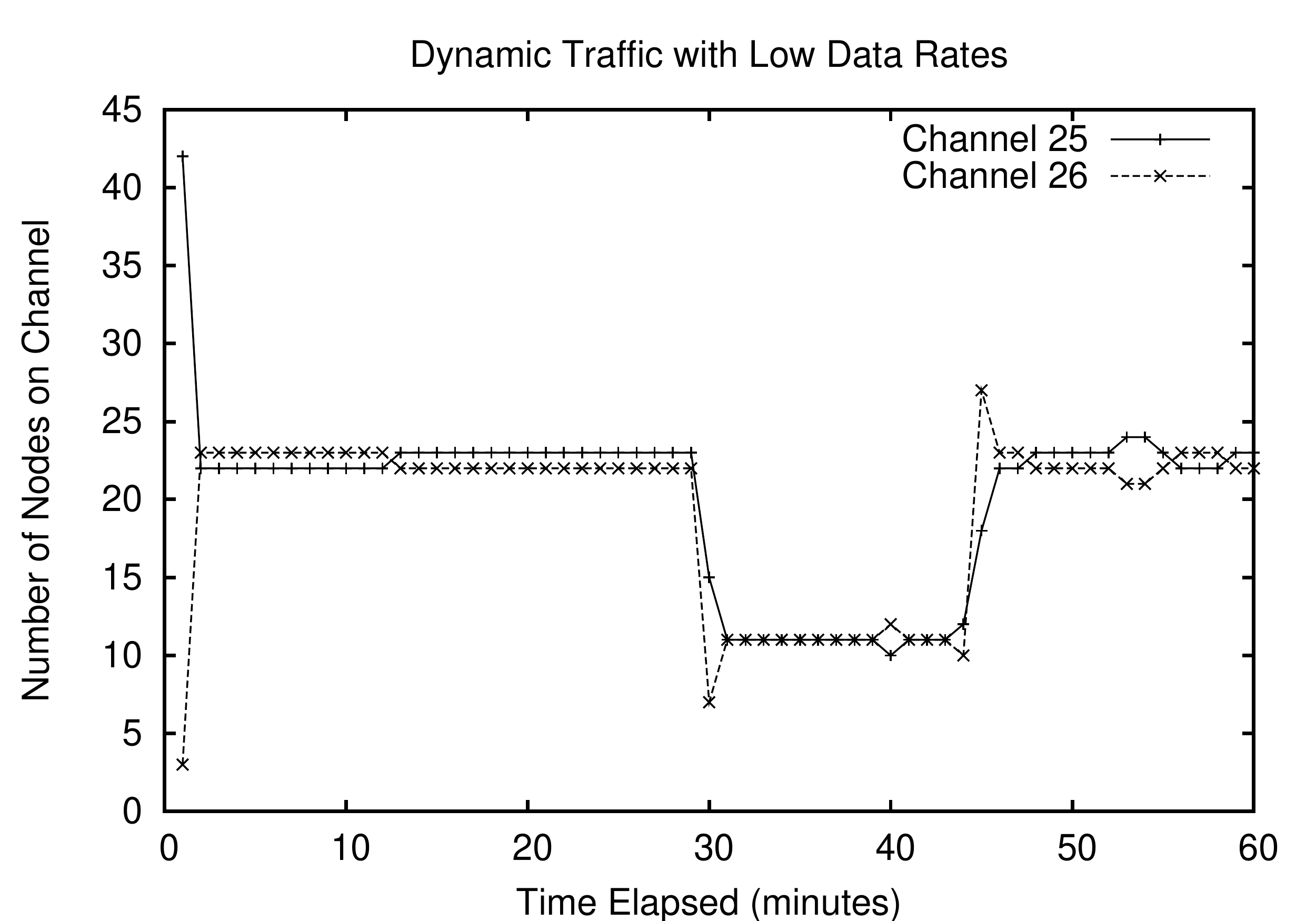}
\includegraphics[width=0.49\linewidth]{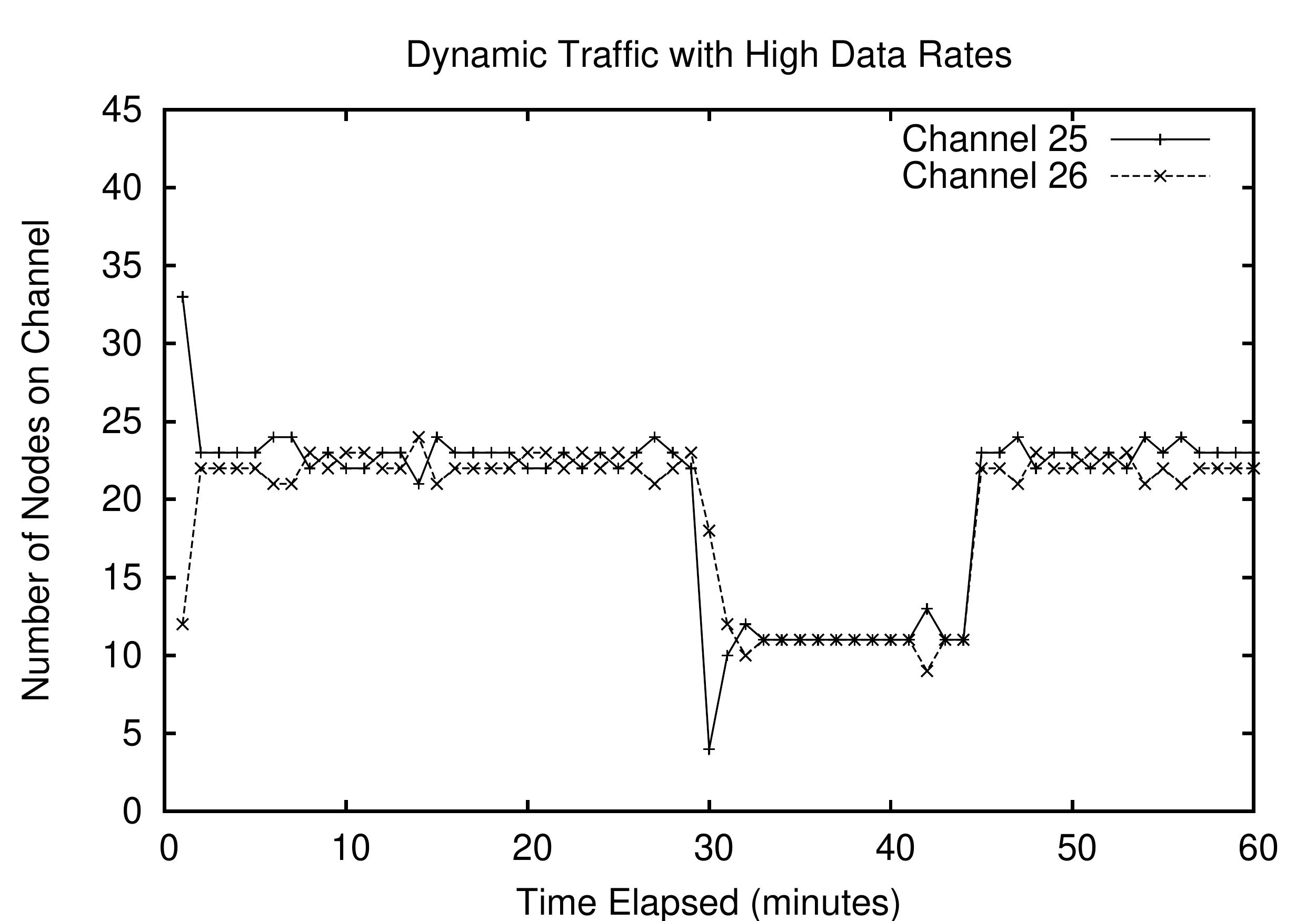}
\caption{Distribution of 45 nodes on each of the two operating
  channels over time with dynamic traffic patterns. The left plot
  shows the case when packets are generated at an interval of 1024 ms
  and the right plot shows the case for traffic with 128 ms interval
  at each source node. To generate dynamic traffic environments, we
  randomly turn off approximately half of the source motes (23 nodes)
  at time $t = 30$ minutes. We can see that the nodes fairly
  re-distribute themselves within a short period of time when a
  significant change in the network traffic occurs. When all nodes
  become active again at time $t = 45$ minutes, the original behavior
  is recovered quickly as well.}
\label{fig:channel-select-dynamic} 
\end{figure}

\begin{figure}[t] \centering
\includegraphics[width=0.49\linewidth]{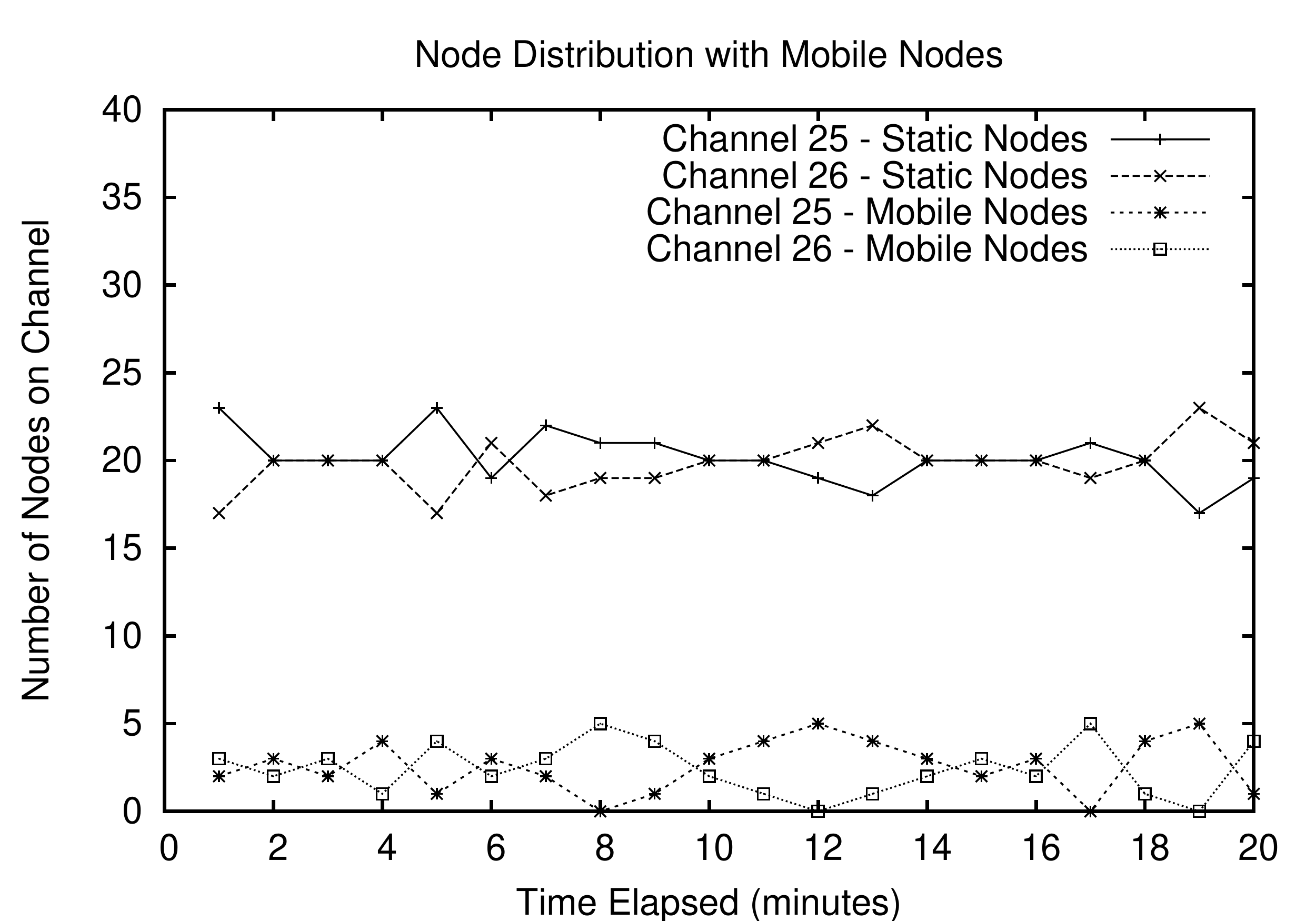}
\caption{Distribution of nodes on each of the two operating channels
  over time with mobile nodes' traffic. 40 nodes stay stationary on
  the testbed and two volunteers each carry two and three nodes
  respectively on the hallway for 20 minutes as the mobile devices.
  All nodes in the experiment transmit one packet each 128 ms.  While
  the distribution of the nodes are not as stable as the results
  observed in the experiments when all nodes are stationary, the
  distribution of the nodes are comparably fair among the two
  channels.}
\label{fig:channel-select-mobile} 
\end{figure}

\begin{figure}[t] \centering
\includegraphics[width=0.49\linewidth]{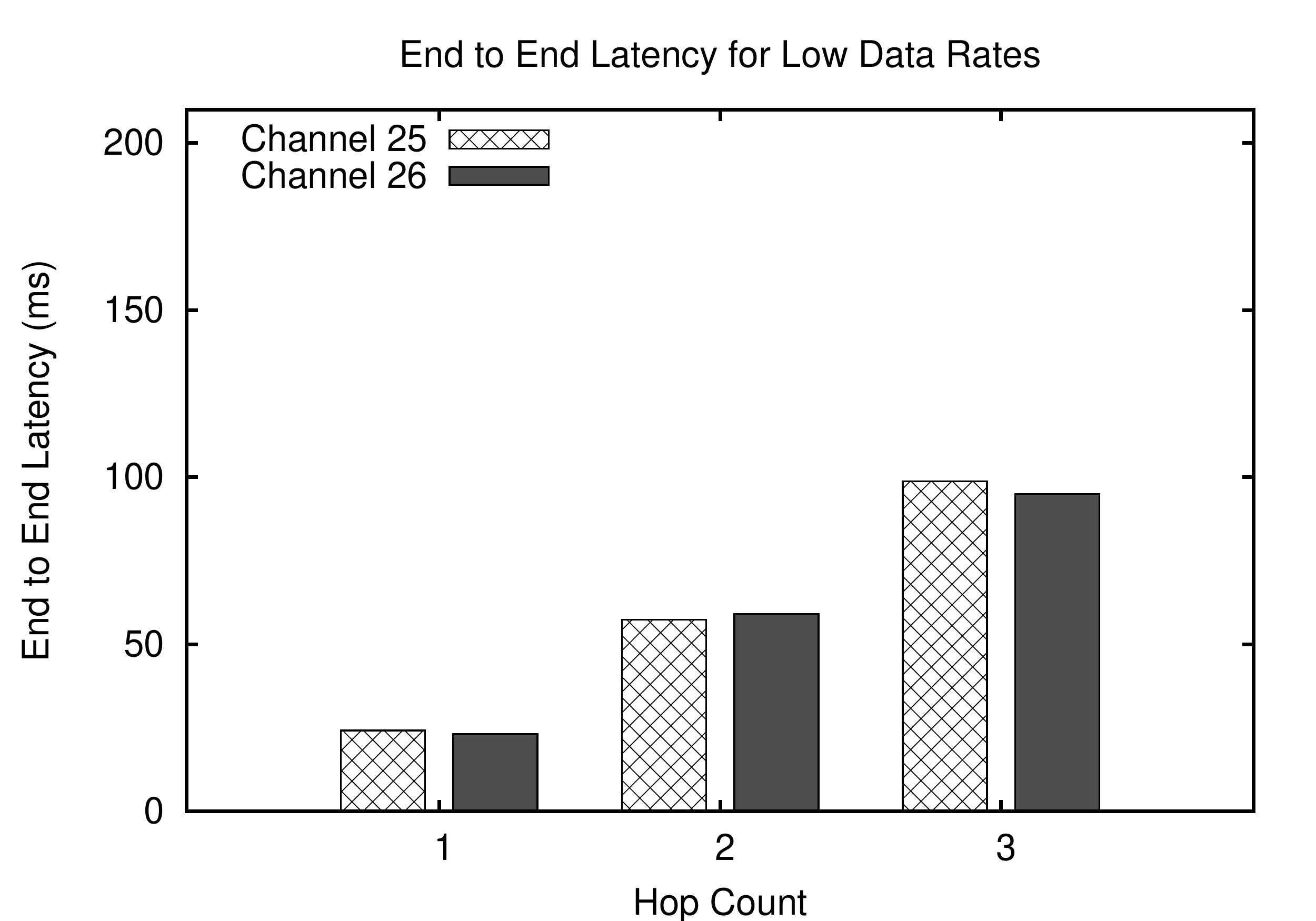}
\caption{End to end latency of packets generated from nodes on
  different channels with steady traffic at a packet interval of 1024
  ms. Nodes connected to different hops on the backbone tree network
  have different average latency values.}
\label{fig:latency-steady-low} 
\end{figure}

\begin{figure}[t] \centering
\includegraphics[width=0.49\linewidth]{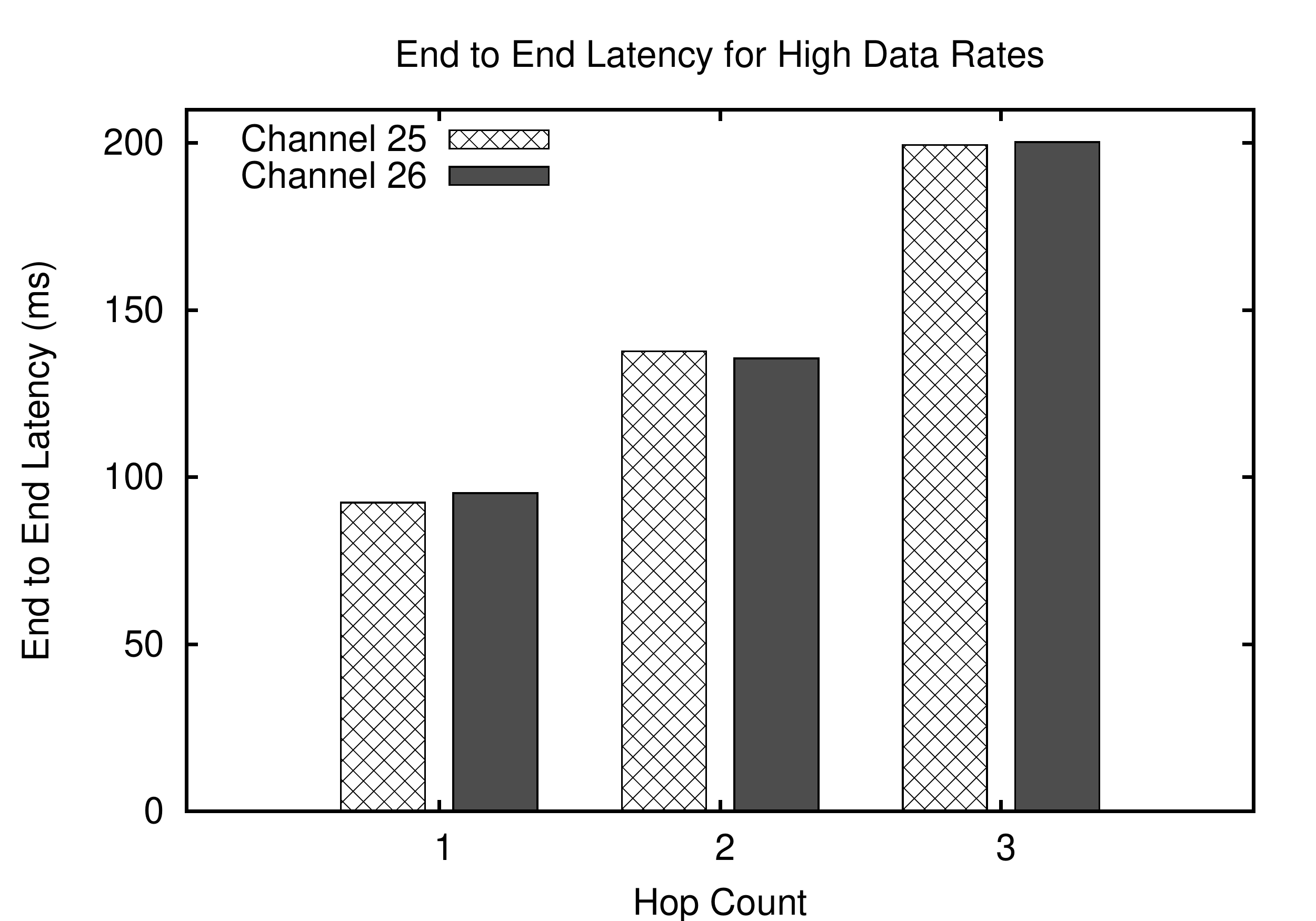}
\caption{End to end latency of packets generated from source nodes on
  different channels with steady traffic at a packet interval of 128
  ms. The higher data rates lead to more packets on each relay node
  and therefore, longer packet latencies compared to the low traffic
  rates presented in Figure~\ref{fig:latency-steady-low}.}
\label{fig:latency-steady-high} 
\end{figure}

\begin{figure}[t] \centering
\includegraphics[width=0.49\linewidth]{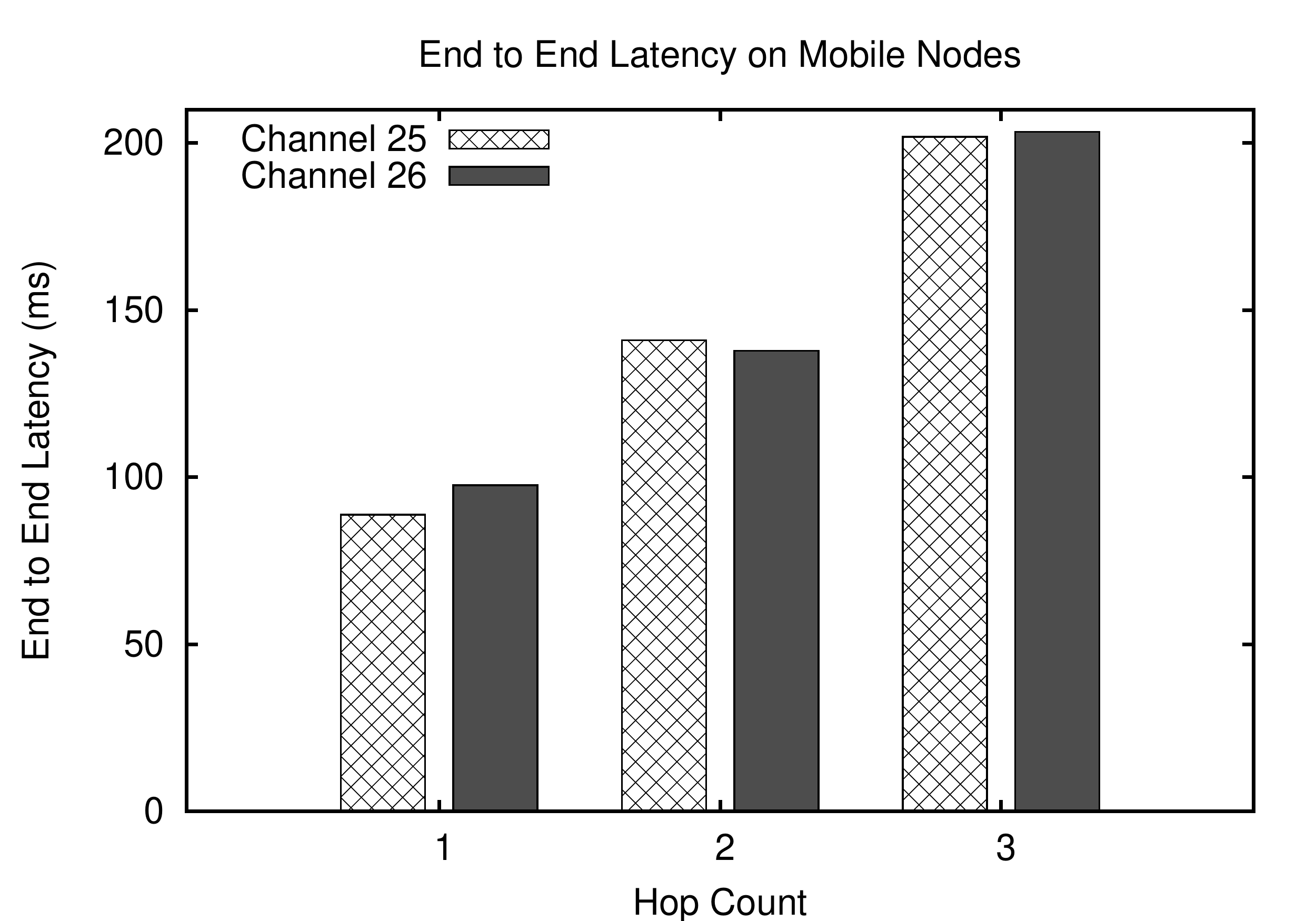}
\caption{End to end latency of packets generated from nodes on
  different channels with mobile nodes. The five mobile nodes and the
  other 40 stationary nodes generate steady traffic with an interval
  of 128 ms. The latency observed at the mobile devices are comparable
  with the latency of the packets generated from stationary nodes (see
  Figure~\ref{fig:latency-steady-low} and
  Figure~\ref{fig:latency-steady-high}).}
\label{fig:latency-steady-mobile} 
\end{figure}

\begin{figure}[t] \centering
\includegraphics[width=0.49\linewidth]{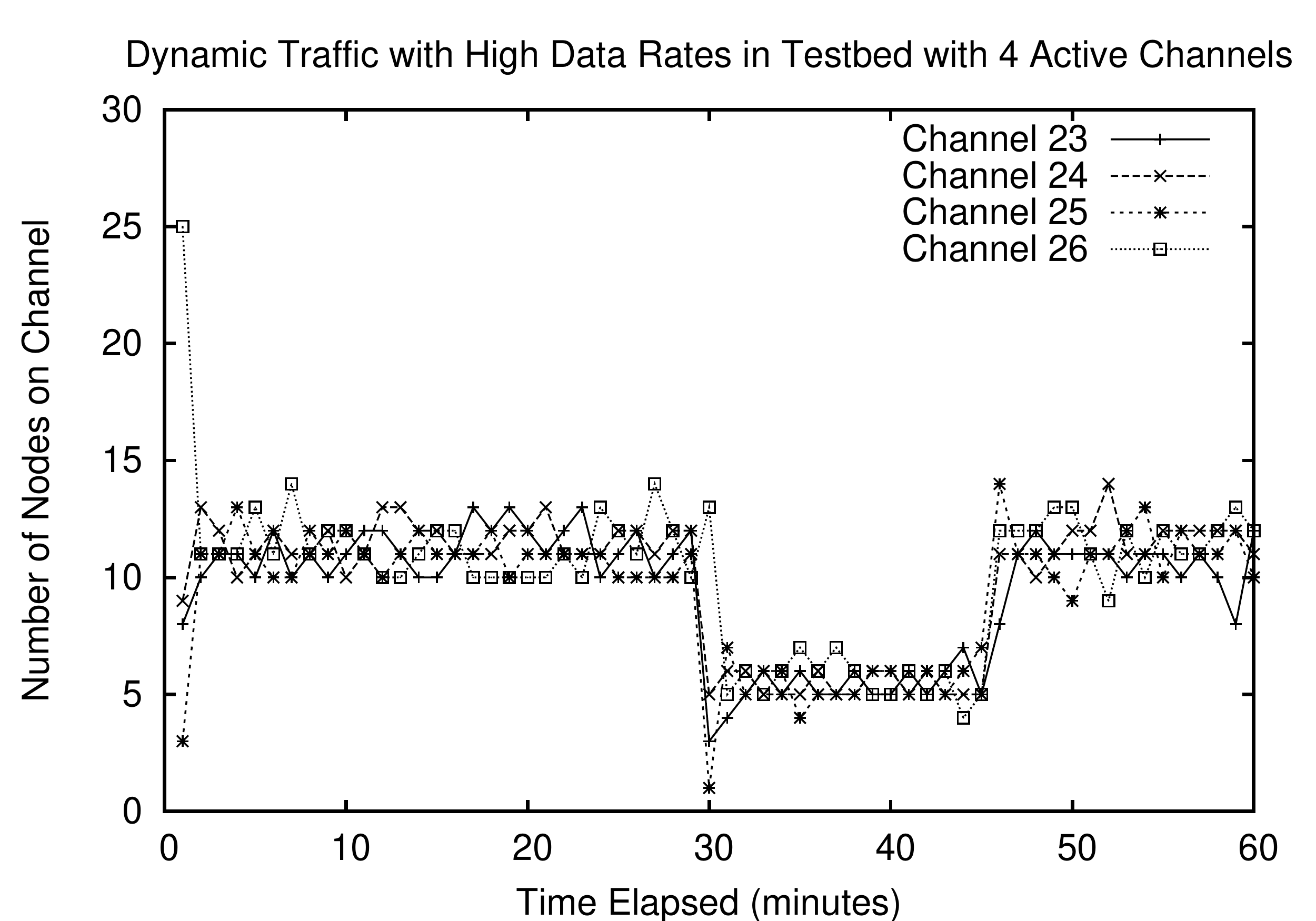}
\includegraphics[width=0.49\linewidth]{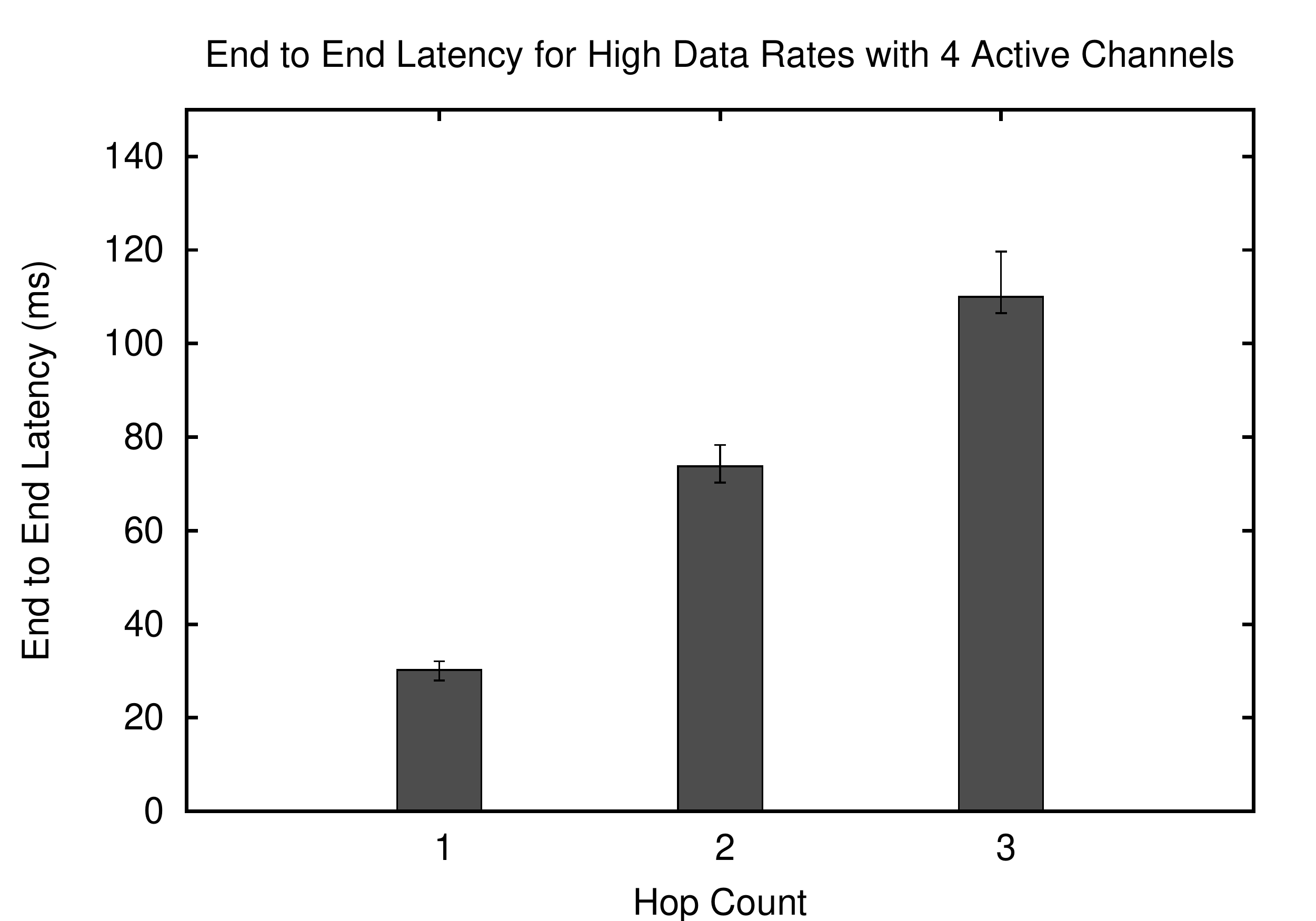}
\caption{Node distribution (left) and latency with respect to the hop
  count of the source node (right). Even when a higher number of
  wireless channels are available for the source nodes, the
  distribution of nodes over multiple channels are fair and the
  latency is fairly distributed over the four different wireless
  channels of our interest. Similar to the results in
  Figure~\ref{fig:channel-select-dynamic}, we experiment with the
  dynamic traffic patterns with high data rates (i.e., half of the
  nodes are switched off at $t = 30$ and turned back on $t = 45$). }
\label{fig:four-channel-experiment} 
\end{figure}

\subsection{Evaluation Environment and Metrics}
\label{sec:eval-env}

As the sections until now discusses, \DynaChanAl~ is a simple and
light weight scheme that can be practically executed on a highly
resource constraint mote-class platform. Furthermore, there is no
existing work that takes the goal of channel allocation for mobile WSN
nodes with the goal of end-to-end latency. Therefore, the goal of this
evaluation is to show that such a simple scheme can effectively
distribute nodes on different channels. Thus we begin the evaluation
to show that this is true (Sections \ref{sec:eval-channel-selection} -
\ref{sec:eval-queuing-latency}). The later part of the evaluation
(Section \ref{sec:eval-throughput}) will confirm that the achieved
throughput of \DynaChanAl~ does not degrade the maximum throughput
that is achievable by our hardware and software platforms in our
target indoors environment. Our evaluations are based on experiments
performed using real mote-class devices deployed in an indoor
testbed. The testbed consists of 45 Tmote Sky motes deployed in
various locations on a single floor hallway (see Figure
\ref{fig:testbed}) which act as the source nodes. We deploy eight
additional Tmote Sky devices to form the backbone network for our two
different channels of interest. The tree network is formed using the
collection tree protocol (CTP) \cite{ctp}, the de-facto standard tree
routing protocol for TinyOS 2.x applications. In our tests we use
Zigbee channels 25 and 26 to minimize the effect of IEEE 802.11 WiFi
interference (see Figure~\ref{fig:channel}). To experiment and monitor
the effect of mobile source devices, two volunteers carry five Tmote
Sky devices at walking speed in the hallway where our testbed and
backbone network is deployed. We note that \DynaChanAl~ is implemented
in TinyOS 2.1.

To quantify the performance of \DynaChanAl~, we define the following
performance metrics. First we take the distribution of nodes on
different channels to see how effectively \DynaChanAl~ distributes
nodes on different wireless channels. Second, we use the end-to-end
latency of the packets generated at the source nodes on each channel.
This second metric is an important metric that indicates whether
\DynaChanAl~ successfully achieves its primary goal of minimizing the
end-to-end latency of all generated packets.  Finally, using the total
network throughput at the root nodes, we show that \DynaChanAl~
effectively uses the available wireless bandwidth.

In our experiments we set parameters as the following. First, for the
$LQI_i$ metric we set $n=1$ during the channel seeking phase and
$n=10$ in the channel monitoring phase. Second, $D_i^q$, as described
in Section~\ref{sec:def_metric}, is computed in run-time on a per
packet basis using packet overhearing techniques. The channel seeking
phase in \DynaChanAl~ is executed every 30 seconds to tolerate dynamic
channel environments. Finally, we set both delay and LQI thresholds,
$\tau_D$ and $\tau_{qli}$ as 90\% (details in Section
\ref{sec:channel-monitoring}) and the application specific end-to-end
delay threshold as 500 ms.

\subsection{Channel Selection of Mobile and Stationary Source Nodes}
\label{sec:eval-channel-selection}

The first set of results that we present aims to show that
\DynaChanAl~ effectively distributes multiple source nodes on the two
different channels that we use in our experiments. We use three types
of traffic patterns and two types of traffic loads for all of our
experiments. We define our three traffic patterns of interest as the
following:

\begin{itemize}
\item{\textit{Steady and Stationary Traffic:} Each source node
    generates packets at a fixed interval with no mobility in the
    environment.}
\item{\textit{Dynamic and Stationary Traffic:} Each source node begins
    by generating steady traffic but a subset of the source nodes are
    switched off and then back on during the experiment. There are no
    mobile source nodes in the experiment.}
\item{\textit{Mobile Traffic:} Each source node generates steady
    traffic throughout the experiment and a subset of the source nodes
    are mobile at human walking speed.}
\end{itemize}

For each of the traffic patterns described above we experiment with
two different traffic loads, \textit{low data rates} (i.e., 1024 ms
packet generation interval at each source node) and \textit{high data
  rates} (i.e., 128 ms packet generation interval).

We first present the results for steady and stationary traffic with
the two types of traffic loads in
Figure~\ref{fig:channel-select-steady}.  Both the left (packet
interval 1024 ms) and right (packet interval 128 ms) plots indicates
that \DynaChanAl~ successfully distributes nodes fairly.  Each point
in the plots represent the average number of nodes in each channel per
minute. The small yet consistent changes in the nodes' channel
distributions are caused by variances introduced in the wireless
channel (e.g., non-deterministic human movement within the testing
environment). In the initial stage of all our experiments, nodes
select their channels based on only a small number of previous samples
(or even none in some cases), and therefore, there is a noticeable
difference in the number of nodes on the two channels.

We generate dynamic traffic patterns by randomly turning off
approximately half of the source nodes (23 nodes) from our testbed.
We make this change 30 minutes after the beginning of the experiment.
These nodes are re-activated after 15 minutes of node blackout (i.e.,
45 minutes in to the experiment). Both left (low data rate) and right
(high data rate) plots of Figure~\ref{fig:channel-select-dynamic} show
that with \DynaChanAl~ nodes can discover and position themselves on
an effective wireless channel regardless of the amount of traffic in
the network. We can notice that the fairness of node distribution on
the two channels break slightly at the beginning of a major traffic
pattern changes (e.g., when nodes are turned off or when they are
suddenly turned on). This fluctuation is relaxed as soon as nodes
notice the changes in traffic conditions and adapt to such changes.
According to Figure~\ref{fig:channel-select-dynamic} this process
takes $ < \sim$1 minute.

Finally, since source nodes can be mobile in applications such as
healthcare WSN systems~\cite{medisn}, we observe how the mobility of
nodes affect the distribution of nodes on multiple wireless channels.
For this, we take five nodes from our testbed to act as mobile nodes
(specified as blue circles in Figure~\ref{fig:testbed}) and generate
mobile traffic.  Two volunteers each carry two and three nodes
respectively for 20 minutes and perform continuous walks on the
hallway shown in Figure~\ref{fig:testbed}.  The experiments are
performed with all source nodes (both mobile and stationary)
generating eight packets each second (i.e., high data rate).  We show
that the number of stationary and mobile nodes on each of our two
channels of interest with respect to experiment duration in
Figure~\ref{fig:channel-select-mobile}.  One can notice that the
source nodes change their wireless channels more frequently when a
subset of the nodes are mobile compared to the case where all the
nodes are stationary (compare with right figure of
Figure~\ref{fig:channel-select-steady}). Such increased amount of
channel switches result from more frequent execution of the channel
seeking phase. We argue that the link quality fluctuations of the
links connecting the mobile nodes with the backbone network causes the
mobile nodes to initiate the channel seeking phase more frequently
(mostly due to the LQI comparison in
Figure~\ref{fig:monitor-diagram}).  Therefore this opens the
possibility of more frequent channel switching. This, in turn,
fluctuates the delay on the backbone network, causing additional
stationary nodes to start the channel seeking phase as well. In any
case, Figure~\ref{fig:channel-select-mobile} indicates that despite
continuously having node mobility in the wireless network,
\DynaChanAl~ successfully distributes nodes on the two channels
effectively.

\subsection{End to End Queuing Latency}
\label{sec:eval-queuing-latency}

During our experiments, we also collect the amount of end-to-end
queuing latency encountered for each packet received at the root of
the tree network. Using these latency measurements we determine if the
delay on the two channels are fairly distributed and minimized.  We
point out that given a specific amount of traffic, the most efficient
way to distribute nodes on multiple channels (when end-to-end latency
of the packets is the core consideration point) is to balance each
channel with the \emph{same} end-to-end latency. Therefore, an optimal
scheme will show even latency on all used wireless channels. We
present the results with steady traffic (explained in
Section~\ref{sec:eval-channel-selection}) for both low and high data
rates in Figures~\ref{fig:latency-steady-low}
and~\ref{fig:latency-steady-high}, respectively. In the figures, we
plot the average end-to-end delay observed by packets generated from
nodes connected to different hops in the backbone tree network.
Results indicate that the latency is (as expected) lower with low data
rates and increases with increasing amount of traffic. Also, as the
number of hops increases, the end-to-end latency grows as well. While
these observations are intuitive, one interesting point to notice from
Figures~\ref{fig:latency-steady-low} and~\ref{fig:latency-steady-high}
is the differences in the increasing slopes of the latency as the
number of hops increase. We notice that while the latency increases
gradually with hop count in the low data rate experiments, a
significant amount of end-to-end latency is seen even at the source
nodes that are connected to the first hop nodes of the backbone
network in the high data rate experiments. We speculate that such
behavior is due to the significant amount of packets that are
``stacked up'' at the relay nodes close to the gateway (i.e., first
hop relay nodes) when the amount of traffic in the network is
significantly high. Such traffic congestion increases the queue size
at specific relay nodes and thus, this increases the packets' queuing
delays. In any case, despite the unavoidable higher end-to-end latency
in such cases, we can observe that the latency is almost equal on the
two active wireless channels. This indicates that \DynaChanAl~
achieves semi-optimal node distribution with respect to minimizing the
end-to-end delay with stationary nodes.

The delay results obtained from the experiment with mobile source
nodes are presented in Figure~\ref{fig:latency-steady-mobile}. Since
the traffic patterns that the source nodes generate are the same as
the high traffic rate experiments (packet interval of 128 ms) the
results look similar to Figure~\ref{fig:latency-steady-high} with only
slightly higher latency values. We conjecture that such higher
latencies despite the same number of source nodes and traffic are
caused by the overhead introduced from the increased number of channel
seeking phases and the fluctuating wireless channel selections. Still
the latency distribution on the two active channels are highly
correlated, thus, \DynaChanAl~ is achieving the minimal possible
latency on both wireless channels.

\begin{table}[t]
\centering
\begin{tabular}{c|c}
\hline
Seeking 2 Active Channels    & 63.76 ms\T\B\\
\hline
Seeking All 16 Possible Channels    & 353.37 ms\T\B\\
\hline
\end{tabular}
\caption{Latency introduced by channel seeking phase with different
  number of channels that are seeked within the phase. The
  measurements are gathered experimentally from our tests.}
\label{tab:seek-latency}
\end{table}

Finally, we note that each channel seeking phase (channel seeking
delay + channel switching delay) took an average 63.76 ms when only
channels 25 and 26 were part of the channel seeking phase. This was
done by pre-configuring the source nodes with the active channels.
When all 16 channels were probed during the channel seeking phase
(i.e., no pre-configuration) the average channel seeking delay was
353.37 ms. These results, summarized in Table~\ref{tab:seek-latency},
show that \DynaChanAl~ indeed effectively selects channels with
minimal channel seeking overhead and these experimental results match
what we expected in the empirical studies shown in
Section~\ref{sec:empirical}.

\subsection{Throughput of \DynaChanAl~ }
\label{sec:eval-throughput}

To see how efficient \DynaChanAl~ performs in terms of end-to-end
network throughput, we compare the amount of data received at the root
node (gateway) in our experiments with the optimal throughput value of
IEEE 802.15.4-based networks. First, while the IEEE 802.15.4
standard~\cite{ieee802.15.4} states that the maximum throughput is 250
Kbps for 2.4 GHz radios, we take an experimental approach to see how
much throughput can \emph{actually} be achieved when using TinyOS 2.1
and Tmote Sky nodes.  For this, we set one receiver and place five
transmitter nodes within a single communications range. The
transmitter nodes sent back-to-back packets (i.e., as fast as
possible) with each packet size being 114 bytes. When a carrier sense
multiple access with collision avoidance (CSMA/CA) based medium access
control (MAC) protocol is fully used (TinyOS 2.1 implements the
B-MAC~\cite{BMAC} as the default MAC protocol), we observed a
throughput of 38.92 Kbps and the maximum throughput was 49.19 Kbps
when we disabled the clear channel assessment (CCA) checks in the
MAC. For a single transmitter with back to back packet transmissions,
the maximum throughput achieved was 42.61 Kbps. This indicates that
the throughput of 38.93 Kbps for multiple nodes is a result of channel
saturation. We will compare the throughput achieved by \DynaChanAl~
with the saturated throughput since the amount of traffic that the
nodes in our experiments generate easily saturates a 802.15.4 wireless
medium. Such low throughput values with the hardware and software
platform that we use (e.g., TelosB motes and TinyOS 2.x) is also
observed in previous work such as
\cite{petrova2006performance,Flush07}. We conjecture that the overhead
introduced by TinyOS' software stack with the speed limitations of the
serial interfaces in the hardware are the main causes of the
performance degradation (e.g., memory copy operations to/from the
radio's buffer, receive pointer reservations until the received
process is terminated from the upper most layer, cross layer
communications, etc.).  These experimental values represent the
maximum possible throughput at a single receiver node when all
transmitting nodes are connected via \emph{singlehop links}. In the
experiments performed for \DynaChanAl~, we incorporate a multihop tree
network of relay nodes which mobile user nodes connect for
transmitting their data to the root of the tree. Such network
architectures can potentially decrease the throughput of the wireless
network due to its multihop nature. In any case, for our testing
environments we observed an average throughput of 33.67 Kbps per
wireless channel.  Given that the gateway issues acknowledgment
packets to all packet receptions and all nodes perform full CSMA/CA
(including CCA checks), this throughput is 86.51\% of the maximum
achievable performance in a saturated environment.  Despite the
multihop network architecture, since the main bottleneck of the
throughput performance is the data flow within TinyOS and the local
device, the performance degradation compared to the optimal case
(maximum singlehop throughput) is not significant.  We conjecture that
the small performance difference that we see is mainly caused by the
communication and computational overhead introduced by the Collection
Tree Protocol (CTP)~\cite{ctp} that we use to form the tree relay
network. Overall, for both channels we observed an aggregate
throughput of 67.34 Kbps.

\subsection{Performance with Increasing Number of Usable Wireless
  Channels}
\label{lab:four-channel}

To confirm that \DynaChanAl~ performs effectively as we predicted in
the previous subsections even with increasing number of usable
wireless channels, we perform a simple experiment on the testbed in
Figure~\ref{fig:testbed} with the backbone infrastructure installed
for four different wireless channels (i.e., Zigbee channels 23, 24, 25
and 26).  In this setting, we generate dynamic/stationary traffic (as
described in Section~\ref{sec:eval-channel-selection}) with high data
rates.  Again at $t = 30$ minutes we turn off 23 nodes and switch them
back on at $t = 45$ minutes to emulate dynamic traffic patterns. We
present the results of the nodes' distribution on multiple channels
(left) and the end-to-end delay with respect to the source node's hop
count with error bars that indicate the minimum and maximum values
(right) in Figure~\ref{fig:four-channel-experiment}. The results
indicate that both the distribution of nodes and the distribution of
end-to-end latency are fair over the four active wireless
channels. These results provide evidence that \DynaChanAl~ is scalable
with increasing number of wireless channels.

\section{Notable Aspects of \DynaChanAl~}
\label{sec::discuss}

\noindent\paragraph{Applicability with Different Protocols:}
\DynaChanAl~ is a fully distributed scheme that is suitable for WSN
systems with multihop connections and incorporate a wireless backbone
architecture. In such networks, systems combine different schemes to
optimize the performance of the WSN system as a whole (i.e., network
protocols, medium access control (MAC) protocols, etc.). Examples of
such WSN systems include~\cite{medisn,codeblue,Wisden04}. We point out
that \DynaChanAl~ is light-weight both in terms of memory usage and
computational power. Therefore, \DynaChanAl~ can be used along with
other algorithms with minimal additional overhead. First, our scheme
is mainly dependent on the packets and decisions of the mobile source
nodes while being less dependent on the infrastructure nodes (i.e.,
only require reply messages). Therefore, \DynaChanAl~ minimally
interacts with the protocols that relay nodes use. Second, the metrics
collected for \DynaChanAl~ can be used to combine with other metrics
to make improved decisions for pre-existing protocols as well (e.g.,
low power MAC protocols that could benefit from minimizing the delay
or routing protocols that can use the latency information to select
latency effective paths).

\noindent\paragraph{Minimizing Energy Consumption:} Finally in terms
of minimizing idle listening times, we show in this work that channel
seeking delays can be minimal (i.e., $< 400 ms$). In \DynaChanAl~ we
can further minimize the idle waiting times by pre-specifying the
active wireless channels. As seen in the results obtained from
Section~\ref{sec:eval-queuing-latency}, seeking only a subset of
active channels instead of all 16 can reduce the seeking delay by up
to 81.97\%. This potentially leaves room for extending the nodes'
lifetime. On the other hand we note that trying to overhear the next
hop's packets to gather the updated LQI and $D_i^q$ values can
increase idle listening at the resource limited source nodes. In such
cases, explicitly transmitting the parameters in a separate data
packet can decrease the total energy consumption, given that for
CC2420 radios the energy spent on idle listening is greater than that
for transmitting. In this case, the lower energy consumption is a
tradeoff with a small amount of communication overhead.  Another way
of minimizing the idle listening time while keeping the communication
overhead low, is synchronizing the source nodes with the next hop
node.  By learning the transmission patterns of the next hop node, the
source node can perform radio duty cycling with respect to this
transmission pattern to conserve energy and overhear the updated
piggybacked parameters simultaneously.

\section{Limitations and Future Work}
\label{sec:future}

The current proposal of \DynaChanAl~ requires a pre-deployed backbone
network on multiple channels which should be carefully deployed with
considerations on the estimated network capacity usage. While this is
a realistic case in applications such as MEDiSN \cite{medisn} where
the positions of the backbone network nodes and the network capacity
usage is carefully monitored and engineered, removing the requirement
of a pre-deployed backbone network on multiple channels would
significantly increase the number of applications that \DynaChanAl~
can support. For this, \DynaChanAl~ should invade other layers of the
network stack including the routing and MAC layers. We should think as
future work what the tradeoffs are in keeping \DynaChanAl~ as an
independent layer (i.e., higher portability to applications but less
optimized) or as part of a system as a whole (i.e., highly optimized
with less portability).

\section{Conclusion}
\label{sec:sum}

In this work we propose and evaluate \DynaChanAl~, a distributed
channel allocation scheme for WSN systems with the goal of minimizing
end-to-end queuing latency on multiple wireless channels. \DynaChanAl~
experimentally measures the queuing delay at each hop of a multihop
network and aggregate these measurements to assure that a mobile
source nodes can make accurate estimates of the expected end-to-end
queuing latency. \DynaChanAl~ also incorporates link quality indicator
(LQI) values to classify the quality of the local wireless channel
environments. Our evaluation is based on experimental results from a
45 node indoor testbed to demonstrate the effectiveness of such a
simple and light-weight scheme in actual mote-class platforms. Our
results show that \DynaChanAl~ effectively distributes multiple nodes
on different wireless channels with steady, dynamic and mobile traffic
patterns. The channel selection of \DynaChanAl~ also assures that the
end-to-end queuing delay is minimal on all active channels, which is
the end goal of our protocol. To our knowledge this work is the first
work to consider the expected end-to-end queuing delay as a main
design metric for wireless channel allocation in WSN systems research.

\bibliographystyle{elsarticle-num}
\bibliography{sensors}
\end{document}